\documentclass[superscriptaddress,showkeys]{revtex4-2}
\usepackage{graphicx}
\usepackage{multirow}
\usepackage{placeins}
\usepackage{dcolumn}
\usepackage{bm}
\usepackage[colorlinks=true, citecolor=red, urlcolor = violet, linkcolor= purple, bookmarks=true]{hyperref}
\usepackage{float}
\usepackage{amsmath,amsfonts,amssymb}
\usepackage{natbib}
\usepackage{hyperref}
\usepackage{breakurl}
\usepackage{xcolor,verbatim,multirow}
\usepackage[normalem]{ulem}
\usepackage{subcaption} 
\usepackage{wrapfig}
\usepackage{adjustbox}
\usepackage{caption}
\usepackage{mathtools}
\usepackage{pgfplots}
\usepackage{mathrsfs}
\usepackage{booktabs}
\begin{document} 

\title[]{Shadows and parameter estimation of rotating quantum corrected black holes and constraints from EHT observation of M87* and Sgr A*}
\author{Heena Ali} \email{Heenasalroo101@gmail.com}
\affiliation{Centre for Theoretical Physics, Jamia Millia Islamia, New Delhi-110025, India }
\author{Shafqat Ul Islam} \email{shafphy@gmail.com}
\affiliation{Astrophysics and Cosmology Research Unit, School of Mathematics, Statistics and Computer Science, University of KwaZulu-Natal, Private Bag X54001,
		Durban 4000, South Africa}
\author{Sushant G. Ghosh}\email{sghosh2@jmi.ac.in}
\affiliation{Centre for Theoretical Physics, Jamia Millia Islamia, New Delhi-110025, India }
\affiliation{Astrophysics and Cosmology Research Unit, School of Mathematics, Statistics and Computer Science, University of KwaZulu-Natal, Private Bag X54001, Durban 4000, South Africa}
\begin{abstract}
The scarcity of quantum gravity (QG) inspired rotating black holes limits the progress of testing QG through Event Horizon Telescope (EHT) observations. The EHT imaged the supermassive black holes, Sgr A* and M87*, revealing an angular shadow diameter of $d_{sh} = 48.7 \pm 7 \mu$as with a black hole mass of $M = 4.0_{-0.6}^{+1.1} \times 10^6 M\odot$ for Sgr A*. For M87*, with a mass of $M = (6.5 \pm 0.7) \times 10^9 M_\odot$, the EHT measured an angular diameter of $\theta_d = 42 \pm 3 \mu$as. We present rotating quantum-corrected black hole (RQCBH) spacetimes with an additional QC parameter $\alpha$ and constrain it by EHT observations. For angular shadow diameter ($d_{sh}$) of Sgr A* at $\theta_o = 50^0$, the bounds are $0.0 \leq \alpha \leq 1.443 M^2$ and $a \in (0, 0.8066 M)$. For $\theta_o = 90^0$, the bounds are $0.0 \leq \alpha \leq 1.447 M^2$ and $a \in (0, 0.894 M)$. While for M87* at inclination $\theta_o = 17^0$, the bounds are $a \in (0, 0.8511 M)$ at $\alpha=0$  and $a \in (0, 0.6157 M)$ at $\alpha=0.8985 M^2$. For $\theta_o = 90^0$, the bounds are $a \in (0, 0.8262 M)$ at $\alpha=0$  and $a \in (0, 0.9799 M)$ at $\alpha=0.4141 M^2$. These results show that $\alpha$ significantly affects the shadows, offering key constraints on QG models. With EHT constraints from Sgr A and M87*, RQCBHs and Kerr black holes are indistinguishable in much of the EHT-constrained parameter space, making RQCBHs strong candidates for astrophysical black holes along with other BHs, e.g., regular black holes and other quantum-corrected solutions.
\end{abstract}

\maketitle

\section{Introduction}\label{Intro}

Black holes, predicted by general relativity (GR) \citep{1915SPAW.......844E}, some of the most enigmatic objects in the universe, serve as profound testing grounds for theories of gravity, particularly in extreme regimes where GR and quantum effects become relevant. Black holes are characterized by an event horizon from which nothing, not even light, can escape \citep{Schwarzschild:1916uq}. Their formation is a natural outcome of GR \citep{Penrose:1964wq}, and they continue to play a key role in efforts to unify GR with quantum physics \citep{Hawking:1976ra}.
The no-hair theorem, a cornerstone of black hole physics within the framework of GR, asserts that the most general, stationary, asymptotically flat, and axially symmetric black hole solution to the Einstein field equations is uniquely described by three parameters: mass $M$, angular momentum $j$, and electric charge $Q$ \citep{Israel:1967wq, Israel:1967za, Carter:1971zc, Hawking:1971vc, Robinson:1975bv}. These parameters are encapsulated in the Kerr-Newman metric \citep{Newman:1965my}, which generalizes the Kerr metric \citep{Kerr:1963ud}—the solution for a rotating, electrically neutral black hole. The no-hair theorem is founded on the uniqueness theorem in GR, which articulates that the Kerr-Newman metric is the only solution to the Einstein-Maxwell equations meeting the conditions of stationarity and axial symmetry. For practical purposes, black holes are typically regarded as electrically neutral since any residual charge rapidly neutralizes through interactions with the surrounding matter \citep{Israel:1967wq, Israel:1967za, Carter:1971zc, Carter:1999mrq, Hawking:1971vc}. Therefore, the Kerr metric often suffices to describe the observed black holes. Ruling out the existence of non-Kerr black holes involves complex observational and theoretical efforts \citep{Ryan:1995wh, Will:2005va}.
The no-hair theorem's mathematical status has been controversial, particularly regarding the assumption of analyticity \citep{Chrusciel:2012jk}. The hypothesis that smooth, analytic metrics accurately depict black holes is essential to the theorem but may not apply in all theoretical frameworks. Modifications to GR, like those in Loop Quantum Gravity (LQG), allow for black holes that deviate from the Kerr solution.
Testing the no-hair theorem concerns investigating potential deviations from Kerr metrics, especially in scenarios proposed by MTG. Such investigations may indicate black holes with additional parameters, as LQG and other MTGs suggested, challenging the traditional no-hair hypothesis.   LQG has emerged as an advantageous theory for handling the singularity problem of GR and has been used recently to effectively \citep{Boehmer:2007ket,Campiglia:2007pb,Gambini:2008dy,Battista:2023iyu}. These singularities pose significant challenges, as they signal a breakdown of the theory and an incapability to describe the fundamental nature of spacetime. LQG explains these singularities by proposing a discrete structure of spacetime at the Planck scale. 

The study of black holes has been revolutionized in recent years, particularly with the groundbreaking detection of gravitational waves by LIGO in 2015.  Meanwhile, the VLBI technique is used by the Event Horizon Telescope (EHT) to acquire the high angular resolution required to image supermassive black holes.  VLBI  at a wavelength of 1.3 mm (230 GHz) using Earth-diameter-scale baselines are necessary to resolve the shadows of the core of M87 (referred to as M87*) and the Galactic center of Sagittarius A (Sgr A*) \citep{1974ApJ...194..265B}, which are the two supermassive black holes with the biggest apparent angular sizes \citep{Johannsen:2012vz}. With the birth of the Event Horizon Telescope (EHT), which produced the first images of black hole shadows in M87*   \citep{EventHorizonTelescope:2019dse, EventHorizonTelescope:2019pgp, EventHorizonTelescope:2019ggy,EventHorizonTelescope:2019jan, EventHorizonTelescope:2019ths, EventHorizonTelescope:2019uob}  and Sgr A* \citep{EventHorizonTelescope:2022exc,EventHorizonTelescope:2022urf,EventHorizonTelescope:2022apq,EventHorizonTelescope:2022wok,EventHorizonTelescope:2022wkp, EventHorizonTelescope:2022xqj}, our ability to test theories of gravity has reached phenomenal levels. The black hole shadow is a dark area that appears against the bright background of the accretion disk and surrounding space. When photons get very close to the black hole, near a region called the photon sphere, they create a dark spot in the sky, known as the shadow, which is surrounded by a bright ring of light called the photon ring 
\citep{Synge:1966okc,Bardeen:1973tla,Luminet:1979nyg,Cunningham:1973tf} on the observer's sky whose outline is marked by gravitationally lensed photons \citep{Johannsen:2010ru}.   Applications of shadow in understanding the near-horizon geometry have sparked extensive research into analyzing shadows of black holes in GR \citep{Falcke:1999pj,Vries2000TheAS,Shen:2005cw,Yumoto:2012kz,Atamurotov:2013sca,Abdujabbarov:2015xqa,Cunha:2018acu,Kumar:2018ple,Afrin:2021ggx,Hioki:2009na,Chen:2023wzv,Li:2024abk} and in modified theories of gravity (MTGs) \citep{Amarilla:2010zq,Amarilla:2011fx,Amarilla:2013sj,Amir:2017slq,Singh:2017vfr,Mizuno:2018lxz,Allahyari:2019jqz,Papnoi:2014aaa,Kumar:2020hgm,Kumar:2020owy,Ghosh:2020spb,Guo:2020zmf,Afrin:2021wlj,Vagnozzi:2022moj,Vagnozzi:2019apd,Afrin:2021imp,Gao:2023mjb,Li:2022eue,Sengo:2022jif,Liu:2024lbi}, including in quantum gravity motivated theories  \citep{Liu:2020ola,Brahma:2020eos,KumarWalia:2022ddq,Islam:2022wck,Afrin:2022ztr, Yang:2022btw}. Since extra dimensions have been introduced in modified frameworks of GR, black hole shadows have also garnered significant attention in higher-dimensional spacetimes \citep{Papnoi:2014aaa,Singh:2017vfr,Amir:2017slq,Singh:2023ops,Vagnozzi:2019apd,Hou:2021okc,Vagnozzi:2022tba,Banerjee:2022jog,Lemos:2024wwi}. Black hole shadows provide a powerful tool for conducting strong-field gravitational tests and, ultimately, for testing the no-hair theorem \citep{Johannsen:2010ru,Baker:2014zba,2014ApJ...784....7B,Khodadi_2020,Khodadi_2021,Glampedakis:2023eek,Afrin:2021imp,Afrin:2021wlj,Afrin:2022ztr,Islam:2022wck}. 

In recent years, studies have explored the implication of quantum corrections for rotating black holes. For instance, QC versions of the Kerr black hole, the classical solution describing rotating black holes, have been proposed and analyzed in various contexts \citep{Liu:2020ola,Brahma:2020eos,KumarWalia:2022ddq,Islam:2022wck,Afrin:2022ztr, Yang:2022btw}. These studies reveal that quantum corrections can lead to significant deviations in the black hole's observable properties, such as the shape and size of the shadow, the deflection of light, and the time delay of signals travelling near the black hole. Such deviations provide a potential means of distinguishing quantum-corrected black holes (QCBHs) from their classical counterparts through observations. 
QCBHs emerge from various theoretical frameworks, including LQG and string theory, which suggest modifications to the classical solutions of Einstein's field equations. These modifications often introduce new parameters that can alter the geometry of black holes, particularly near the event horizon, where quantum effects are expected to be most significant.
Lewandowski {\it et al.} \citep{Lewandowski:2022zce} recently introduced a QCBH model in  LQG that modifies the Schwarzschild black hole and resolves the singularity problem by halting collapse at the Planck scale \citep{Lewandowski:2022zce}. Studies have explored QCBH properties like its shadow, photon ring, quasinormal modes and gravitational lensing \citep{Yang:2022btw,Gong:2023ghh,Ye:2023qks,Zhao:2024elr,Liu:2024soc}. 
The dearth of rotating black hole models within LQG has significantly impeded efforts to test LQG theory through observational data. It motivated us to explore rotating or axisymmetric generalizations of the spherical QCBH metric \citep{Lewandowski:2022zce}. The rotating quantum corrected black hole (RQCBH) metric, with an additional parameter $\alpha$, is analogous to the Kerr metric and is derived using the revised Newman-Janis algorithm (NJA). The RQCBH provides a framework for testing LQG theories with astrophysical observations by offering a model incorporating rotation. Moreover, the revised NJA has proven effective in generating rotating metrics from nonrotating seed metrics within LQG models \citep{Liu:2020ola, Brahma:2020eos, Chen:2022nix}. Our research aims to enhance the observational testing of LQG and contribute to a deeper understanding of black hole physics in MTGs.  For  RQCBHs, parameter $\alpha$ can lead to observable deviations in the black hole's shadow properties and gravitational lensing effects, which can be tested against high-precision observational data from the EHT.
This paper investigates the RQCBH model and its implications for the EHT observations of M87* and Sgr A*, focusing on how parameter $\alpha$ can modify the black hole's shadow and the constraints that can be derived from the EHT data. By comparing the projections of the QC model with the observed shadows, we aim to constrain the parameter $\alpha$, offering a potential test for quantum gravity theories. We evaluate the feasibility of constraining $\alpha$ using the shadow observations of M87* and Sgr A* from the EHT.
 The EHT Collaboration provided the first horizon-scale image of M87* in 2019 \citep{EventHorizonTelescope:2019dse, EventHorizonTelescope:2019pgp, EventHorizonTelescope:2019ggy,EventHorizonTelescope:2019jan, EventHorizonTelescope:2019ths, EventHorizonTelescope:2019uob}. With M87* at a distance of $d=16.8$ Mpc and an estimated mass of $M=(6.5 \pm 0.7) \times 10^9 M_\odot$ \citep{EventHorizonTelescope:2019dse, EventHorizonTelescope:2019pgp, EventHorizonTelescope:2019ggy}, the EHT results constrain the compact emission region size to an angular diameter of $\theta_d=42 \pm 3, \mu$as and a circularity deviation $\Delta C \lesssim 0.1$. The shadow of M87* is consistent with predictions for a Kerr black hole according to GR. However, uncertainties in spin measurements and deviations in quadrupole moments leave room for potential modifications to Kerr black holes \citep{EventHorizonTelescope:2019dse, EventHorizonTelescope:2019pgp, EventHorizonTelescope:2019ggy, Cardoso:2019rvt}. In 2022, the EHT released shadow observations of the black hole Sgr A* in the Milky Way, revealing a shadow angular diameter of $d_{sh}=48.7 \pm 7, \mu$as and a thick emission ring with a diameter of $\theta_d=51.8 \pm 2.3 \mu$as. With Sgr A* having a mass of $M = 4.0_{-0.6}^{+1.1} \times 10^6 M\odot$ and a distance of 8 kpc from Earth, these observations align with the expected appearance of a Kerr black hole \citep{EventHorizonTelescope:2022exc,EventHorizonTelescope:2022urf, EventHorizonTelescope:2022apq, EventHorizonTelescope:2022wok, EventHorizonTelescope:2022wkp, EventHorizonTelescope:2022xqj}. The consistency of these results with GR predictions for M87* demonstrates agreement across three orders of magnitude in central mass \citep{EventHorizonTelescope:2022wkp}.

The structure of this paper is as follows: We begin in Sec.~\ref{Sec2} with the construction of the RQCBH metric and discuss generic features of the black hole, including horizon structures and ergoregions. In Sec.~\ref{Sec3}, we use spacetime isometries to infer the RQCBH's Komar mass and angular momentum. Sec.~\ref{Sec4} focuses on the null geodesics and black hole shadows. We also discuss how the QC parameter $\alpha$ affects their shape and size. In Sec.~\ref{Sec5}, we provide the shadow characterization observables and utilize them to estimate the parameters related to RQCBHs. The QC parameter $\alpha$ constraints are derived from the EHT shadow bounds of M87* and Sgr A* for the inclination angles of 17\textdegree~and 50\textdegree~ respectively, and also for 90\textdegree~ for both in Sect.~\ref{Sec6}. Finally, Sect.~\ref{Sec7} provides our conclusions and discusses our findings' implications for future observational QG tests.

\section{Quantum Corrected Black Holes}\label{Sec2}
The metric of the  static and spherical QCBH derived by Lewandowski {\it et al.} \cite{Lewandowski:2022zce}
reads 
\begin{align}\label{metric1}
ds^2=-&\left[1-\frac{2M}{r}+\frac{\alpha M^2}{r^4}\right]dt^2 +
\frac{1}{\left[1-\frac{2M}{r}+\frac{\alpha M^2}{r^4}\right]}dr^2 \nonumber \\ + & r^2(d\theta^2+\sin^2{\theta}d\phi^2).
\end{align}
Here the parameter $\alpha=16\sqrt{3}\pi\gamma^3 $  is Quantum Correction (QC) parameter  with $M$ denoting the black hole mass which coincides with the Arnowitt-Deser-Misner (ADM) mass \cite{Kelly:2020uwj, Parvizi:2021ekr}. We use the natural units, $c = G = \hbar = 1$.
The metric of spherical QCBH encompasses the Schwarzschild black hole when $\alpha$ = 0. Further, the metric (\ref{metric1}) is asymptotically flat.
It is convenient to introduce the parameter   $0<\beta<1$ by 
\begin{equation}
G^2M^2= \frac{4 \beta ^4}{\left(1-\beta ^2\right)^3} \alpha.
\end{equation}
It turns out that for $0<\beta<1/2$, i.e., when 
\begin{equation}\label{Min}
M < M_{\rm min}:=\frac{4}{3\sqrt{3}G}\sqrt{\alpha},
\end{equation}
the metric function $g^{rr}$ has no real root, implying that the metric (\ref{metric1}) does not admit any horizon.  The global causal structure of the maximally extended spacetime is the same as that of the Minkowski spacetime.  Hence the value 
\begin{equation}\label{eq:Min1}
M_{\rm min} = \frac{16\gamma\sqrt{\pi\gamma}}{3\sqrt[4]{3}}\frac{1}{G}
\end{equation}
is a lower bound for BHs produced by our  models   \cite{Zhang:2021wex,Giesel:2021dug,Husain:2021ojz}. The minimal mass is of the order of Planck mass whose value depends on the value of Barbero-Immirzi parameter $\gamma\approx0.2375$ \cite{Lewandowski:2022zce, Ye:2023qks, Meissner:2004ju,Domagala:2004jt}.
Consider the case  of 
$M>M_{\rm min}$, i.e.,  $1/2<\beta<1$, the function $g^{rr}$ has exactly two roots 
$$r_\pm=\frac{\beta\left(1\pm \sqrt{2\beta-1}\right)}{\sqrt{(1+\beta)(1-\beta)^3}}\sqrt{\alpha},$$
where ${r_\pm}$ correspond to event $(r_+)$ and cauchy $(r_-)$  horizons. 
 
Observational evidence, such as the images of M87* and Sgr A* produced by the EHT has confirmed that these black holes are indeed rotating, aligning with the Kerr solution of general relativity \citep{EventHorizonTelescope:2019dse,EventHorizonTelescope:2022wkp}. 
The rotating black hole metrics, such as the Kerr metric, introduce unique physical features like the ergosphere, frame-dragging, and the Penrose process, all of which have significant implications for understanding high-energy astrophysical phenomena, including jet formation and energy extraction from black holes \citep{Penrose:1971uk}.

\begin{figure}[hbt!]
\includegraphics[scale=0.8]{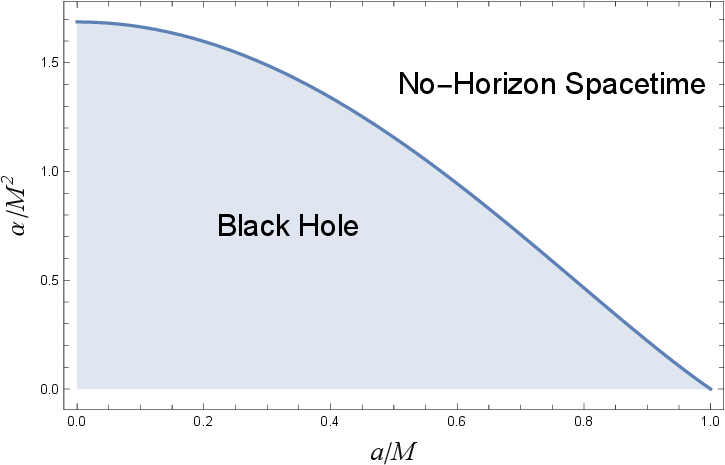}
\caption{\label{fig1}The parameter space ($a$, $\alpha$) for the RQCBH metric. The parameters on the blue curve correspond to extremal RQCBHs with degenerate horizons. The blue curve separates the black hole from no-horizon spacetime.}
\end{figure}

 \paragraph{Rotating black holes:}
 The EHT collaboration unveiled the first images of the black holes M87* and Sgr A*, supporting the presence of rotating black holes consistent with the Kerr solution of GR \citep{EventHorizonTelescope:2019dse, EventHorizonTelescope:2022wkp}, thus indicate that both M87* and Sgr A* are indeed rotating \citep{EventHorizonTelescope:2022xqj}. 
Motivated by this, we seek stationary and axisymmetric (or rotating) counterparts of the spherical QCBH (\ref{metric1}) governed by parameters $M$, $a$ and $\alpha$. The NJA is a powerful method to derive rotating black hole solutions from a static spherically symmetric seed metric without directly solving the Einstein field equation \citep{Newman:1965tw}; the NJA provides a way to generate the Kerr solution from the Schwarzschild metric through a complex coordinate transformation \citep{Newman:1965tw}. Although initially developed in the context of GR, the NJA or modified NJA has been successfully extended to various MTGs \citep{Johannsen:2011dh,Ghosh:2014pba,Kumar:2017qws,Kumar:2019pjp, Kumar:2020hgm,Kumar:2020owy, Ghosh:2021clx,Ghosh:2023nkr}. 
The modified NJA \citep{Azreg-Ainou:2014pra,Azreg-Ainou:2014aqa} has been employed to derive the rotating black hole metric within LQG \citep{Brahma:2020eos, Kumar:2022vfg}. 
Starting with a static, spherically symmetric black hole metric (\ref{metric1}), the application of the revised NJA \citep{Azreg-Ainou:2014pra,Azreg-Ainou:2014aqa} yields the RQCBH metric, which can always be expressed in Boyer-Lindquist coordinates as 
\begin{align}\label{metric}
  ds^{2}= & -\left[\frac{\Delta-a^2 \sin^2\theta}{\Sigma}\right]dt^{2}-2a\sin^2\theta \left[1-\frac{\Delta-a^2 \sin^2\theta}{\Sigma}\right] \nonumber\\
  & \times \; dtd\phi +\sin^2\theta \left[\Sigma+a^2\sin^2\theta(2-\frac{\Delta-a^2 \sin^2\theta}{\Sigma})\right] d\phi^{2} +\frac{\Sigma}{\Delta}dr^{2} +\Sigma d\theta^{2},
\end{align}
with 
\begin{eqnarray}
  \Delta=r^{2}+a^{2}-2M(r)r~~~~~~ M(r)=  M - \frac{\alpha M^2}{2r^{3}}  ~~~~~~~~~~~~~~\Sigma=r^{2}+a^{2}\cos^{2}\theta.  
\end{eqnarray}
Here $a$ is the angular momentum and $\alpha$ is the QC parameter that measures the potential deviation of the metric (\ref{metric}) from the standard Kerr black hole metric. For definiteness, the metric (\ref{metric}) is termed the RQCBH, which encompasses the Kerr black hole as a special case when $\alpha=0$ and reverts to the QCBH metric (\ref{metric1}) when $a = 0$ \cite{Lewandowski:2022zce}.  We depict the parameter space $(a, \alpha)$ in Fig.~\ref{fig1}. The shaded region corresponds to a black hole with two horizons, while for all parameter values ($\alpha_E$, $a_E$) along the blue solid lines, the radius $r_+=r_-$ degenerates, indicating the presence of an extremal black hole. In contrast, the white region represents spacetimes without horizons. 
Apart from the revised NJA, we have also explored the gravitational decoupling (GD) approach as an alternative framework to derive RQCBHs as detailed in \citep{Contreras:2021yxe}. In particular, the GD approach allows us to systematically generate solutions with primary hair and explore nontrivial modifications to the Kerr and Kerr-Newman metrics while ensuring that the deformed metrics, including one obtained here, should satisfy the gravitational field equations.  

\begin{figure}[hbt!]
\includegraphics[scale=0.8]{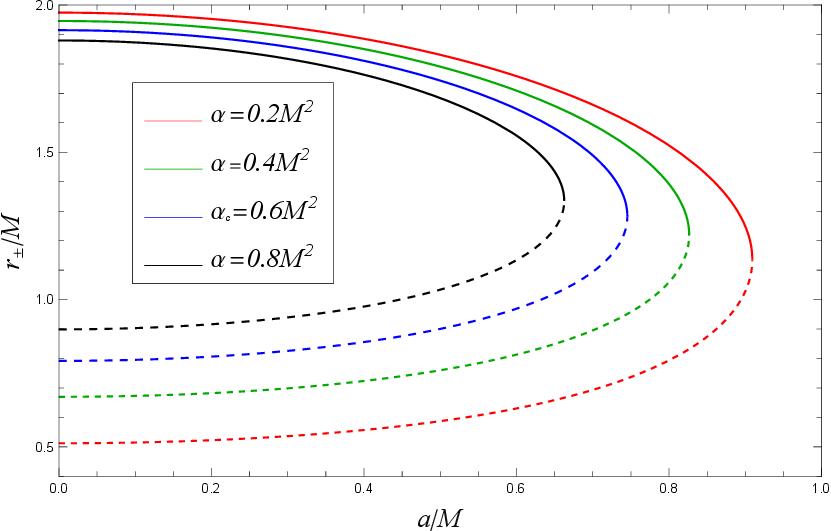}
\caption{\label{fig3}Event (solid curves) and Cauchy (dashed curves) horizons of the RCQBHs for various values of QC parameter $\alpha$ .}
\end{figure}

\begin{figure*}[hbt!]
\centering
\begin{tabular}{c c}
    \includegraphics[scale=0.64]{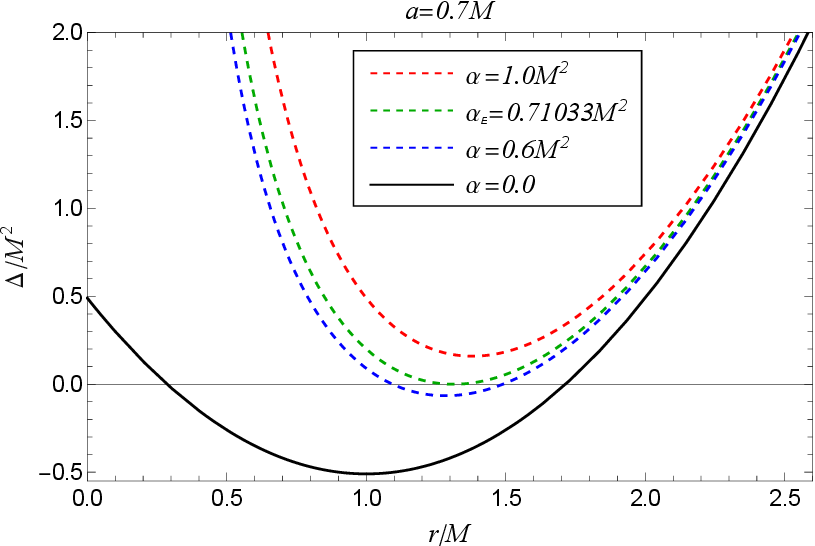}&
    \includegraphics[scale=0.64]{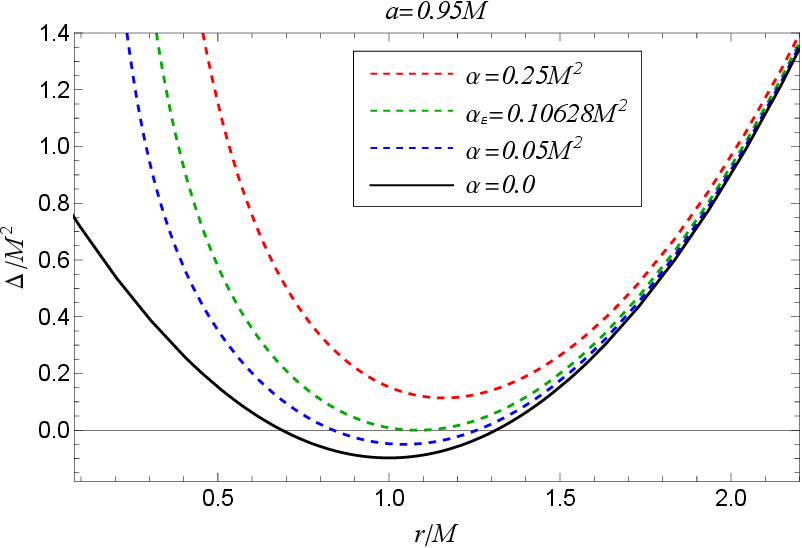}\\
    \includegraphics[scale=0.64]{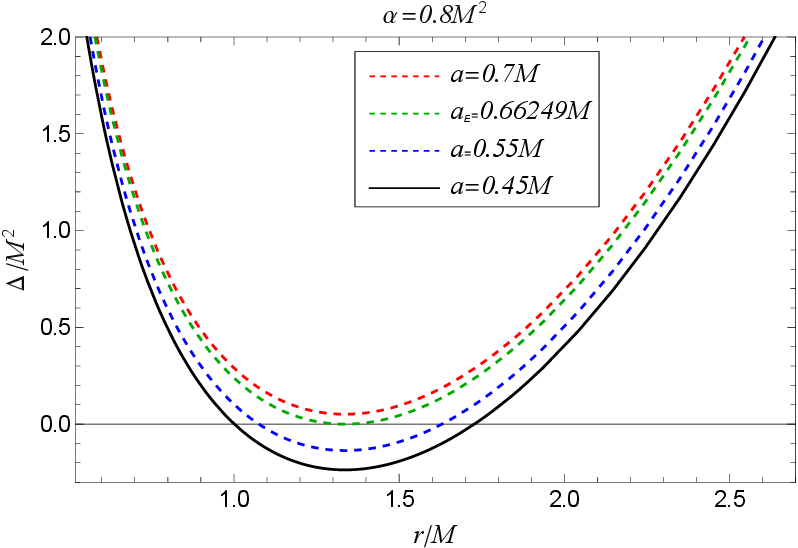}&
    \includegraphics[scale=0.64]{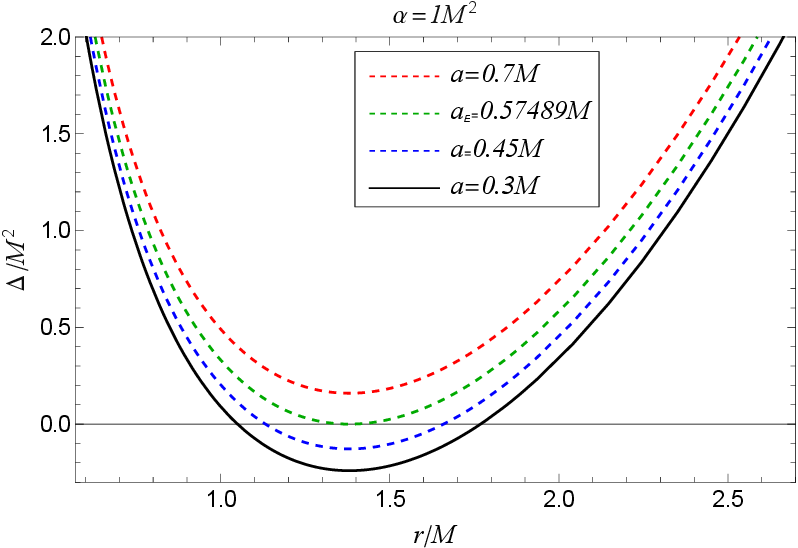}
    \end{tabular}
\caption{Plots showing the variation of $\Delta(r)$ vs $r$ (horizons) for the RQCBH with varying  $\alpha$ (top) and $a$ (bottom) and compared with the Kerr black holes ($\alpha = 0$). The green curves correspond to extremal RQCBHs.}\label{horizons}
\end{figure*}

The metric (\ref{metric}) is singular where $\Sigma\neq0$ and $g^{\alpha\beta}\partial_{\alpha}r\partial_{\beta}r=g^{rr}=\Delta=0$, which admits a maximum of two distinct real positive roots ($r_{\pm}$), equal or no real roots, depending on the values of $a$ and $\alpha$, where $r_+$ denotes the outer (event) horizon and $r_-$ represents the inner (Cauchy) horizon (cf. Fig.\ref{fig3}). When $\alpha=0$, $\Delta=0$ yields
\begin{eqnarray} r_{\pm}^{k} = M\pm\sqrt{M^2-a^2}, 
\end{eqnarray} 
where $r_{\pm}^{k}$ are the horizons associated with the Kerr black hole when $a \leq M$.
Clearly, the metric (\ref{metric}) describes a non-extremal black hole when $r_+ > r_-$, and an extremal black hole when $r_+ = r_-$.
The horizon structure of RQCBH in Fig.~\ref{horizons} shows that for given values of $\alpha$, there exists a critical extremal value $a_E$ of $a$, and similarly, a critical extremal value $\alpha_E$ of $\alpha$ for a given value of $a$, where $\Delta=0$ has a double root. The QCBH exists for $a<a_E$, while a no-horizon arises for $a>a_E$. For instance, for $\alpha = 0.8 M^2$ and $\alpha = 1 M^2$, the values of $a_E$ are 0.66249M and 0.57489M, respectively. Furthermore, $\alpha<\alpha_E$ corresponds to RQCBHs with both Cauchy and event horizons, whereas $\alpha>\alpha_E$ leads to the presence of a no-horizon spacetime. Notably, for $a_E = 0.7M$ and $0.95M$, the corresponding values of $\alpha_E$ are $0.71033 M^2$ and $0.10628 M^2$, respectively. Indeed, $a_E$ decreases as $\alpha$ increases, and similarly, $\alpha_E$ decreases with an increase in $a$.

On the other hand, at the static limit surface (SLS), the asymptotic time-translational Killing vector $\chi^{i}=(\frac{\partial}{\partial t})^{i}$, becomes null, i.e., $\chi^{i}\chi_{i}=g_{tt}=0.$

\begin{figure*}[hbt!]
\centering
\begin{tabular}{c c c c}
    \includegraphics[scale=0.5]{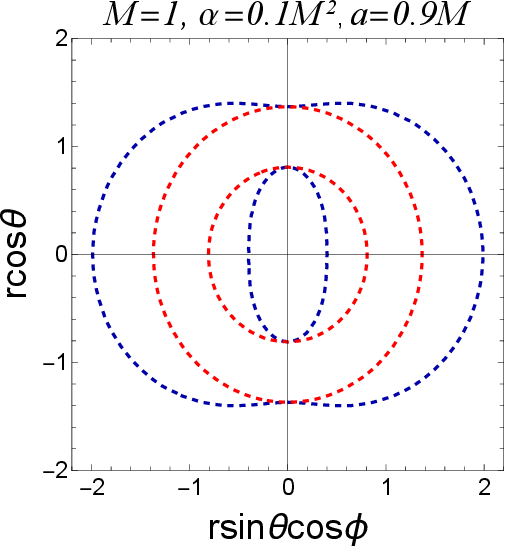}&
    \includegraphics[scale=0.5]{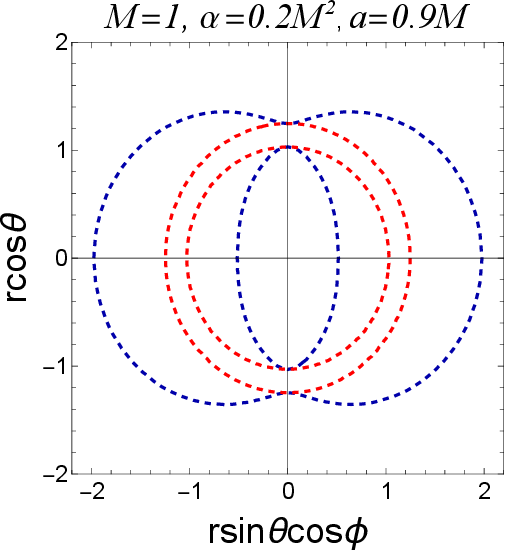}&
    \includegraphics[scale=0.5]{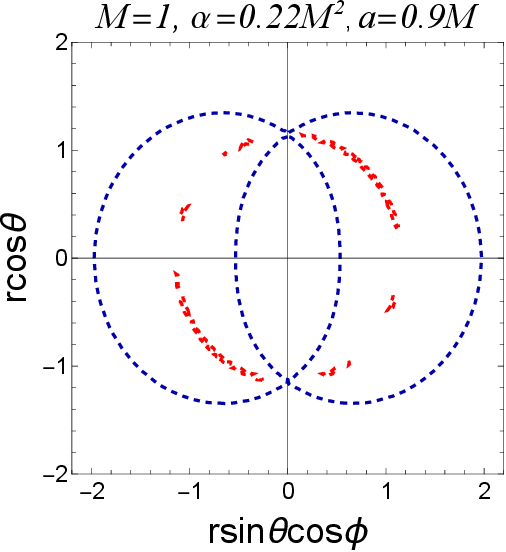}&
    \includegraphics[scale=0.5]{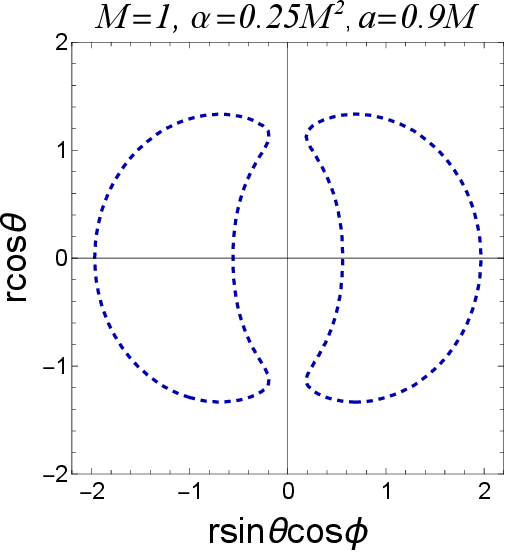}\\
    \includegraphics[scale=0.5]{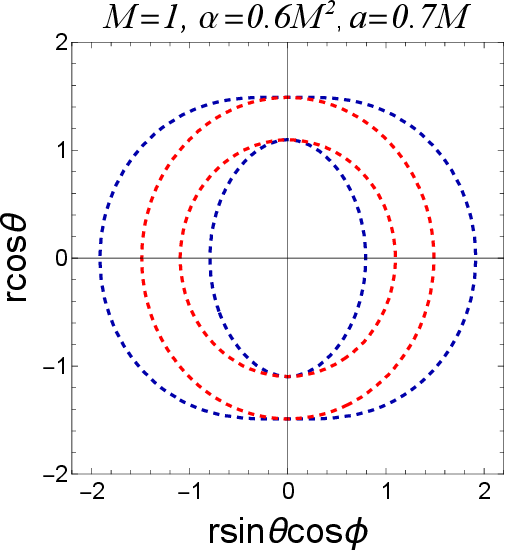}&
    \includegraphics[scale=0.5]{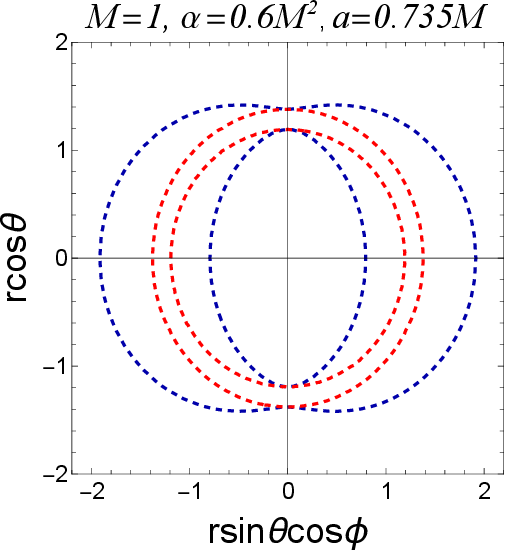}&
    \includegraphics[scale=0.5]{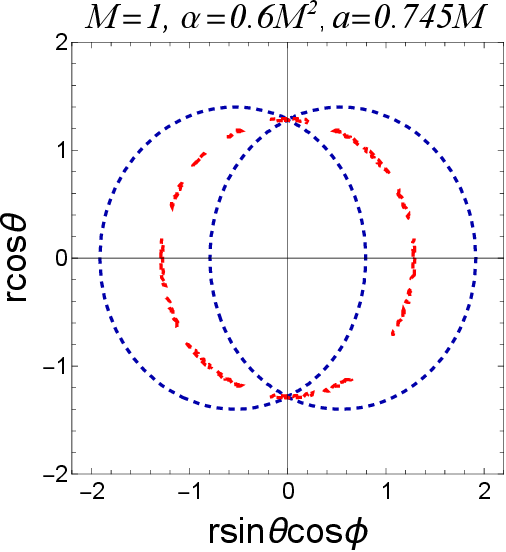}&
    \includegraphics[scale=0.5]{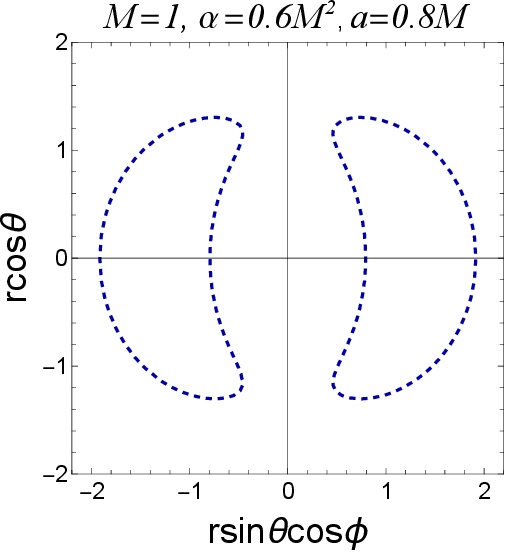}
    \end{tabular}
\caption{The cross-section of the event horizon (outer red curve), SLS (outer blue dotted curve) and ergoregion of the RQCBHs for different values of QC parameter $\alpha$. The increased QC parameter $\alpha$ (top) and rotation parameter $a$ (bottom) lead to disconnected event horizons. }\label{ergo}
\end{figure*}

The larger of the two roots corresponds to the outer SLS, denoted by $r_{SLS}^{+}$. The ergoregion bounded between $r_{+} < r < r_{SLS}^{+}$, where the timelike killing vector $\chi^{i}$ becomes spacelike, and an observer necessarily follows the worldline of $\chi^{i}$, has been shown in Fig.~\ref{ergo} for the RQCBH metric (\ref{metric}). The ergoregions of RQCBHs get larger with increase in $\alpha$ for fixed values of the parameter  $a$, hence are larger than those for the Kerr black hole ($\alpha=0$).
The increase in the parameter $\alpha$ ultimately leads to disconnected event horizons (cf. Fig.~\ref{ergo}). The ergoregion is important owing to the possibility of extracting energy from this region via the Penrose process \citep{Penrose:1971uk}.

\paragraph{Energy Extraction:} To quantify the enhancement in energy extraction efficiency due to quantum corrections, we analyze the impact of an enlarged ergoregion on the Penrose process. The Penrose process, proposed by Penrose \cite{Penrose:1969pc}, allows energy extraction from a rotating black hole due to the presence of an ergosphere. Within this region, particles can follow timelike or null trajectories with negative energy. When a particle splits inside the ergosphere, one fragment may acquire negative energy and fall into the black hole, while the other escapes to infinity with increased energy, effectively drawing energy from the black hole’s rotation\cite{1972ApJ...178..347B,1986ApJ...307...38P,Christodoulou:1970wf,1971NPhS..229..177P}. This work discusses the Penrose process in the context of rotating quantum-corrected black holes. In the classical Kerr black hole, the maximum efficiency of the Penrose process is approximately 20.7\% when extracting rotational energy \citep{Christodoulou:1970wf,Bardeen:1972fi}. However, spacetime geometry modifications alter the ergoregion's location and size in quantum-corrected black holes. The ergoregion extends further than the Kerr case for the RQCBH due to additional metric corrections characterized by the quantum parameter $\alpha$. It directly affects the negative energy orbits within the ergosphere, increasing the potential efficiency of energy extraction. A perturbative analysis suggests that the maximum efficiency $\eta$ can be approximated as 
\begin{equation}
\eta_{\text{max}} \approx 20.7\% + \delta\eta(\alpha),
\end{equation}
where $\delta\eta(\alpha)$ represents the correction due to quantum effects.
For small quantum corrections, previous studies on modified Kerr-like spacetimes indicate that $\delta\eta(\alpha)$ can contribute an increase of a few \%, potentially pushing efficiency to 23-25\% for reasonable values of $\alpha$ \cite{Toshmatov:2019qih,Ghosh:2013ona}. It suggests that quantum corrections can lead to more efficient energy extraction than the classical Kerr scenario. Observationally, this enhancement may have implications for high-energy astrophysical processes such as relativistic jets and gamma-ray bursts, as these  mechanisms are believed to be powered by energy extraction from spinning black holes \cite{Blandford:1977ab,Komissarov:2005wj}. Future studies involving numerical simulations of quantum-corrected black hole geometries could provide more precise estimates of the energy extraction efficiency and its observational signatures.

The frame-dragging effect of the RQCBH becomes evident from the off-diagonal elements of metric (\ref{metric}) viz., $g_{t\phi}$. Due to this effect, a stationary observer outside the event horizon, moving with zero angular momentum concerning an observer at spatial infinity, can rotate with the black hole with an angular velocity given by \citep{Pugliese:2018hju}
\begin{eqnarray}\label{omega1}
   \tilde{\omega}=\frac{d\phi}{d t}=-\frac{g_{t\phi}}{g_{\phi\phi}}=\frac{2ar(M-\frac{\alpha M^2}{2r^3})}{(r^2+a^2)^2-a^2 \Delta\sin^2{\theta}},
\end{eqnarray}
which monotonically increases as the observer approaches the black hole and, ultimately at the event horizon, where the observer begins co-rotating with the black hole, takes the maximum value:
\begin{eqnarray}\label{omega2}
   \Omega= \tilde{\omega} |_{r=r_{+}}=\frac{2ar_+(M-\frac{\alpha M^2}{2r_+^3})}{(r_+^2+a^2)^2}.
\end{eqnarray}
Here $\omega$ is the black hole angular velocity, which in the limit $\alpha=0$ reads
\begin{eqnarray}
    \Omega_{Kerr}=\frac{a}{r_{+}^2+a^2}
\end{eqnarray}
and corresponds to the angular velocity of Kerr black hole \citep{Frolov:2014dta}. It is clear that each horizon point has the same angular velocity (as measured at infinity), and in this sense, the surface of the black hole is rotating \citep{Frolov:2014dta}.

\section{Komar Mass and Angular Momentum}\label{Sec3}
 The Komar formalism provides a conserved definition of mass and angular momentum for stationary, axisymmetric spacetimes, such as rotating black holes. The Komar mass is associated with the timelike Killing vector, representing the system's energy, while the Komar angular momentum is linked to the axial Killing vector and rotation \cite{Komar:1958wp,Wald:1984rg}.
These quantities form the basis for the first law of black hole mechanics, linking changes in mass, angular momentum, and horizon area \citep{Bardeen:1973gs}. The EHT observations provide further reason for a precise description of mass and angular momentum, as these quantities are directly tied to the black hole's observable shadow. 

Adopting the Komar \cite{Komar:1958wp} formalism for defining conserved quantities, we focus on a spacelike hypersurface $\Sigma_t$ extending from the event horizon to spatial infinity. This surface, characterized by constant $t$, has a unit normal vector $n_{\mu}$ \cite{Chandrasekhar:1985kt,Wald:1984rg}. The two-boundary $S_t$ of $\Sigma_t$ is a surface of constant $t$ and $r$, with the outward-pointing unit normal vector $\sigma_{\mu}$. The effective mass of the quantum-corrected black hole can then be expressed as \cite{Komar:1958wp}
\begin{equation}
M_{\text{eff}}=-\frac{1}{8\pi}\int_{S_t}\nabla^{\mu}\eta^{\nu}_{(t)}dS_{\mu\nu},\label{mass}
\end{equation}
where $dS_{\mu\nu}=-2n_{[\mu}\sigma_{\nu]}\sqrt{h}d^2\theta$ is the surface element of $S_t$, $h$ is the determinant of ($2\times 2$) metric on $S_t$, and 
\begin{equation}
n_{\mu}=-\frac{\delta^{t}_{\mu}}{|g^{tt}|^{1/2}},\qquad \sigma_{\mu}=\frac{\delta^{r}_{\mu}}{|g^{rr}|^{1/2}},
\end{equation}
are, respectively, timelike and spacelike unit outward normal vectors. 
The EHT observations offer key insights into quantum gravitational effects near the event horizon by modelling the shadows and estimating parameters like quantum corrections and spin \cite{EventHorizonTelescope:2019dse,EventHorizonTelescope:2022wkp}. 
Thus, mass integral Eq.~(\ref{mass}) turned into an integral over closed 2-surface at infinity:
\begin{align}
M_{\text{eff}}=&\frac{1}{4\pi}\int_{0}^{2\phi}\int_{0}^{\phi}\frac{\sqrt{g_{\theta\theta}g_{\phi\phi}}}{|g^{tt}g^{rr}|^{1/2}}\nabla^{t}\eta^{r}_{(t)}d\theta d\phi\nonumber\\
=& \frac{1}{4\pi}\int_{0}^{2\phi}\int_{0}^{\phi}\frac{\sqrt{g_{\theta\theta}g_{\phi\phi}}}{|g^{tt}g^{rr}|^{1/2}}\left(g^{tt}\Gamma^{r}_{tt}+g^{t\phi}\Gamma^{r}_{t\phi} \right)d\theta d\phi.
\end{align}
To illustrate the procedure, we begin with calculating the effective mass for the simpler case of the Kerr-Newman black hole with  the electric charge $Q$.

\begin{equation}
M_{\text{eff}}=M-\frac{Q^2}{2r^2a}\left[ (r^2+a^2)\tan^{-1}\left(\frac{a}{r}\right)+a r\right]\label{mass0},
\end{equation}
which is the same as in the \cite{Kumar:2020hgm,Modak:2010fn}. Next, using the metric~(\ref{metric}), we first calculate the effective mass of the RQCBH, and it reads
\begin{equation}
M_{\text{eff}}=M-\frac{3\alpha M^2}{2r^4}\left[\frac{ar}{3}+\left(a+r^2\right)\tan^{-1}\left(\frac{a}{r}\right)\right]\label{massRQCBH2},
\end{equation}
which is corrected due to the QC parameter $\alpha$. 
Clearly, the equations ~(\ref{mass0}) and ~(\ref{massRQCBH2}) reduce to effective mass, $M$ of the Kerr black hole when $Q=0$ and $\alpha=0$ respectively.

\begin{figure*}
	 \begin{tabular}{c c}
    \includegraphics[scale=0.65]{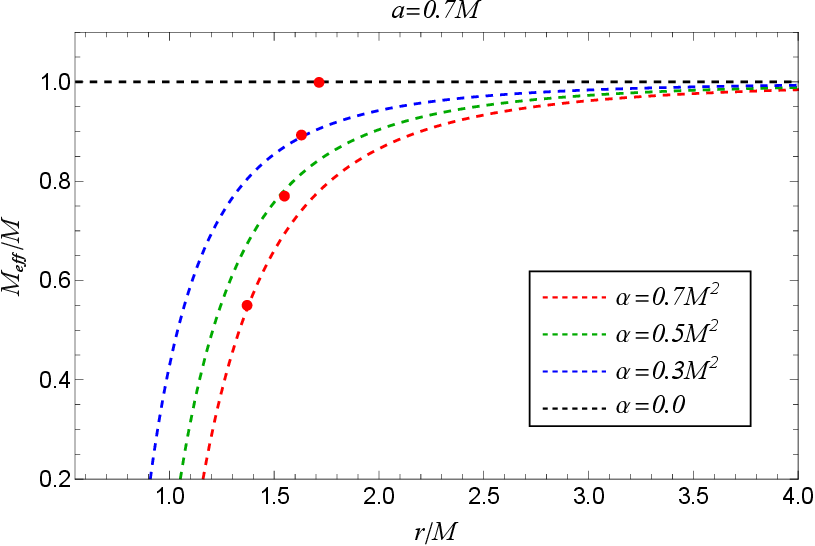}&    
    \includegraphics[scale=0.65]{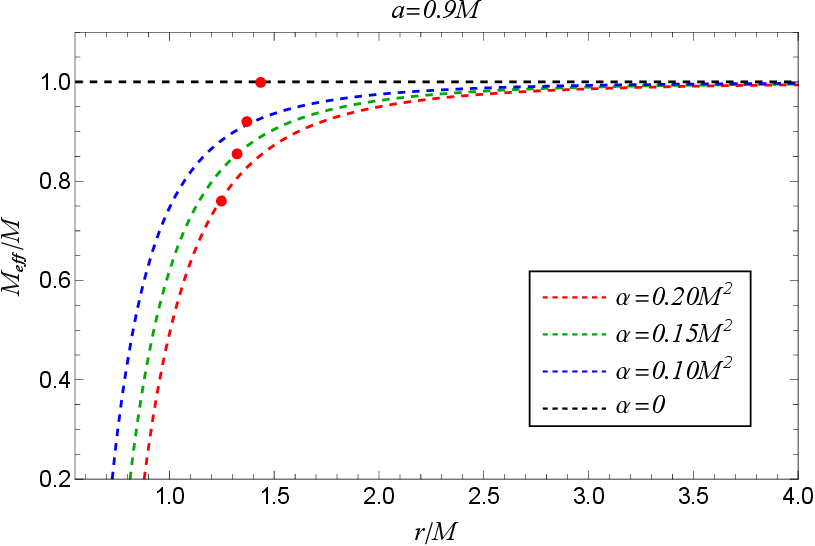}\\
    \includegraphics[scale=0.65]{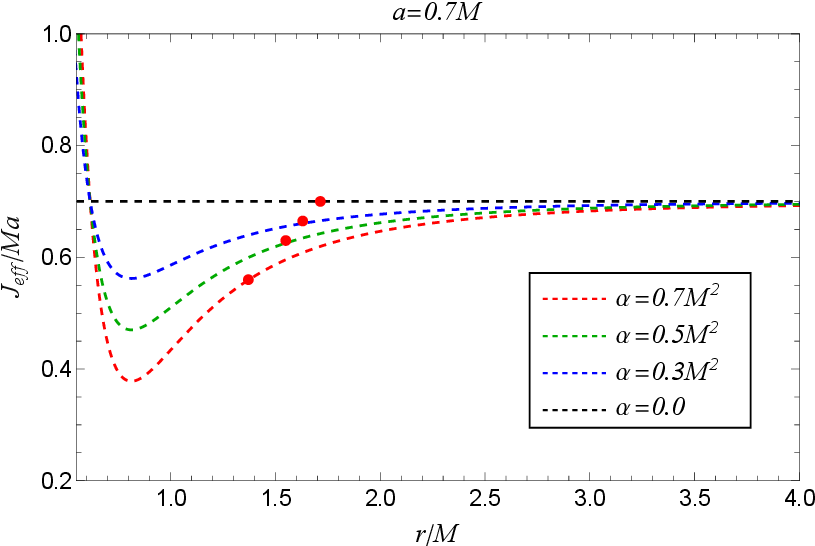}&
    \includegraphics[scale=0.65]{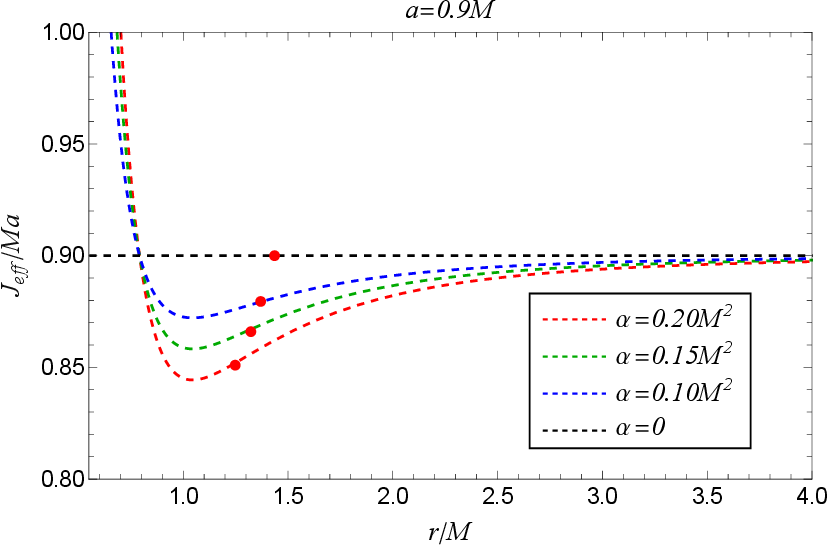}
   \end{tabular}

 	\caption{The behaviour of effective mass $(M_{\text{eff}}/M)$ and angular momentum $(J_{\text{eff}}/Ma)$ vs $r$ for different values of the black hole parameters. Black curves correspond to the Kerr black hole, and red dots in each curve denote the location of the event horizon. }
 \label{Komar}
\end{figure*}
Now, we use the spacelike Killing vector $\eta^{\mu}_{(\phi)}$ to calculate the effective angular momentum  \cite{Komar:1958wp}
\begin{equation}
J_{\text{eff}}=\frac{1}{16\pi}\int_{S_t}\nabla^{\mu}\eta^{\nu}_{(\phi)}dS_{\mu\nu}.\label{ang}
\end{equation}
Using the definitions of the surface element, Eq.~(\ref{ang}) recast as
\begin{align}
J_{\text{eff}}=&-\frac{1}{8\pi}\int_{0}^{2\phi}\int_{0}^{\phi}\nabla^{\mu}\eta^{\nu}_{(t)}n_{\mu}\sigma_{\nu}\sqrt{h}d\theta d\phi\nonumber\\
=& \frac{1}{8\pi}\int_{0}^{2\phi}\int_{0}^{\phi}\frac{\sqrt{g_{\theta\theta}g_{\phi\phi}}}{|g^{tt}g^{rr}|^{1/2}}\left(g^{tt}\Gamma^{r}_{t\phi}+g^{t\phi}\Gamma^{r}_{\phi\phi} \right)d\theta d\phi.
\end{align}

The effective angular momentum for the Kerr-Newman black hole is given by 
\begin{equation}
J_{\text{eff}}=Ma+\frac{(4Mr-Q^2)a^2+r^2Q^2}{4ar}-\frac{Q^2(a^2+r^2)\tan^{-1} \! \left(\frac{a}{r}\right)}{4a^2r^2}\label{jj},
\end{equation}
which is the same as obtained earlier \citep{Modak:2010fn}.
We calculate the effective angular momentum for the RQCBH as
\begin{eqnarray}
J_{\text{eff}}&=&Ma+\frac{3\alpha M^2}{4a^2r^4}\left[ar^3+\frac{a^3r}{3}+\left(a^4-2a^2r^2-r^4\right) \tan^{-1} \! \left(\frac{a}{r}\right)\right].\label{jjj}
\end{eqnarray}

The effective angular momentum equations~(\ref{jj}) and~(\ref{jjj}) restore the value $J_{\text{eff}}=Ma$, which is the value for the Kerr black hole, in the asymptotic limit $r\to\infty$. As a result, the QC parameter's effects become less pronounced as one travels farther from the black hole.
Fig.~\ref{Komar} illustrates how normalized effective mass and angular momentum fluctuate with radial distance $r$ for different black hole characteristics. The normalized $M_{\text{eff}}$ values asymptotically approach unity for large $r$. For fixed values of $a$, it is evident that when the field parameter $\alpha$ increases, the values of $M_{\text{eff}}/M$ and $J_{\text{eff}}/Ma$ decrease. 
It is commonly recognized that the stationary black hole horizon is not generated by the Killing vectors  $\eta^{\mu}_{(t)}$ or $\eta^{\mu}_{(\phi)}$; rather, it is their particular linear combination \cite{Chandrasekhar:1985kt,Kumar:2020hgm} as
\begin{equation}
\chi^{\mu}=\eta^{\mu}_{(t)}+\Omega \eta^{\mu}_{(\phi)},
\end{equation}
such that, although it is a Killing vector only at the horizon, $\chi^{\mu}$ is globally timelike outside the event horizon \cite{Chandrasekhar:1985kt,Kumar:2020hgm}. At the event horizon connected to $\chi^{\mu}$, the Komar conserved quantity reads as \cite{Komar:1958wp}
\begin{eqnarray}
K_{\chi}&=&-\frac{1}{8\pi}\int_{S_t}\nabla^{\mu}\chi^{\nu}dS_{\mu\nu} .\nonumber\\\label{kx}.
\end{eqnarray}
Using the definition of surface element and metric (\ref{metric}), Eq.~(\ref{kx}) yields
\begin{eqnarray}
K_{\chi}&=&M_{\text{eff}}-2\Omega J_{\text{eff}},\nonumber\\
&=&\frac{-M}{{r_+}^3\left(a^2+{r_+}^2\right)}\left[2M\alpha {r_+}^2+Ma^2\alpha-{r_+}^5a^3{r_+}^3\right].\label{ST}
\end{eqnarray}

Next, we analyse the thermodynamic quantity associated with the RQCBH metric (\ref{metric}). The black holes are characterized by their mass ($M_+$ ),
which can be expressed in terms of horizon radius ($r_+$ ) and obtained by solving $\Delta(r_+)=0$ for $M_+$,  which gives
$$M_+=\frac{r_+}{\alpha}\left({r_+}^2+\sqrt{{r_+}^4-a^2\alpha-\alpha {r_+}^2}\right) .$$
 We compute the black hole horizon temperature to comprehend the significance of the above-conserved quantities \cite{Chandrasekhar:1985kt}.  Since a black hole behaves as a thermodynamic system, its temperature $T$ can be determined from its surface gravity $\kappa$ as
\begin{eqnarray}
T_+&=&\frac{\kappa}{2\pi}=\frac{\Delta'(r_+)}{4\pi(r_+^2+a^2)},\nonumber\\
&=& \frac{1}{2\pi\left(r_+^2+a^2\right)}\left[\left(r_+-M\right)-\alpha\frac{M^2}{r^3}\right],\label{temp}
\end{eqnarray}
whereas entropy is defined as follows:
\begin{equation}\label{entropy}
S_+=\frac{A}{4}=\pi(r_+^2+a^2).
\end{equation}
\begin{equation}\label{entropy1}
S_+T_+=\frac{\Delta'(r_+)}{4}=\frac{1}{2\alpha r}\left[2\alpha r^2+a^2\alpha-3\sqrt{{r_+}^4-a^2\alpha-\alpha {r_+}^2}\right].
\end{equation}
Equations (\ref{ST})$-$(\ref{entropy}) clearly infer that
\begin{equation}
K_{\chi}=M_{\text{eff}}-2\Omega J_{\text{eff}}=2S_+T_+.
\end{equation}
Consequently, the Smarr formula is satisfied given that the Komar conserved quantity, which corresponds to the null Killing vector at the event horizon $\chi^{\mu}$, is twice the product of the black hole entropy and the horizon temperature \cite{Smarr:1972kt,Bardeen:1973gs}.

\section{Null Geodesics around Rotating Black Holes and black Hole Shadows}\label{Sec4}
 Light rays from background sources with an impact parameter exceeding the critical value are strongly deflected. They can reach an observer, while those with smaller impact parameters plunge into the event horizon, forming a dark region known as the \emph{shadow}, surrounded by a bright photon ring \citep{Synge:1966okc,Bardeen:1973tla,Luminet:1979nyg,Cunningham:1973tf}. Synge (\citeyear{Synge:1966okc}) and Luminet (\citeyear{Luminet:1979nyg}) derived a formula for the angular radius of the photon region around a Schwarzschild black hole.  Bardeen (\citeyear{Bardeen:1973tla}) subsequently studied the shadow of a Kerr black hole, showing that its spin distorts the shadow shape. The photon ring, which outlines the shadow, is determined by the spacetime geometry, making its shape and size key for extracting black hole parameters \citep{Afrin:2021imp,Kumar:2018ple} and uncovering near-horizon gravitational features. Extensive analytical and numerical work has since explored shadows of various black hole types \citep{Vries2000TheAS,Shen:2005cw,Amarilla:2010zq,Yumoto:2012kz,Amarilla:2013sj,Atamurotov:2013sca,Abdujabbarov:2016hnw,Abdujabbarov:2015xqa,Cunha:2018acu,Mizuno:2018lxz,Mishra:2019trb,Shaikh:2019fpu,Kumar:2020yem,Afrin:2021imp,Afrin:2024khy,Islam:2022wck,KumarWalia:2022aop,Ghosh:2022kit,Afrin:2021wlj,Atamurotov:2024nre,Molla:2024lpt}, including for parameter estimation \citep{Afrin:2024khy,Vachher:2024fxs,Afrin:2021imp,Kumar:2018ple} and testing gravity theories \citep{Kramer:2004hd}.
To find the null geodesics followed by photons in the RQCBH spacetime (Eq.~\ref{metric}), we start from the Hamilton-Jacobi equation \citep{Carter:1968rr,Chandrasekhar:1985kt}: 
\begin{eqnarray} \label{HmaJam} \frac{\partial S}{\partial \lambda} = -\frac{1}{2}g^{\alpha\beta}\frac{\partial S}{\partial x^\alpha}\frac{\partial S}{\partial x^\beta}, \end{eqnarray} 
where $\lambda$ is the affine parameter along the geodesics, and S is the Jacobi action. The time translational and rotational symmetry of Eq.~(\ref{metric}) ensures conserved quantities for null geodesics: energy $\mathcal{E} = -p_t$ and axial angular momentum $\mathcal{L} = p_{\phi}$, where $p_{\mu}$ is the photon's four-momentum—additionally, the Petrov type D character of the metric guarantees the existence of Carter's separable constant. Thus, the Jacobi action can be expressed as: \begin{eqnarray} S = -\mathcal{E} t + \mathcal{L} \phi + S_r(r) + S_\theta(\theta), \end{eqnarray} where $S_r(r)$ and $S_\theta(\theta)$ are functions of the radial and angular coordinates, respectively. 

The geodesic motion of photons around the black hole is necessary for the shadow formation. To study the same, we analyse the motion of a test particle in stationary and axially symmetric spacetime. Since metric (\ref{metric}) is independent of $t$ and $\phi$, these are the cyclic coordinates with corresponding killing vectors  $\chi_{(t)}^{\mu}=\delta _t^{\mu }$  and $\chi_{(\phi)}^{\mu}=\delta _{\phi }^{\mu }$, whose existence further implies that the corresponding four-momentum components $p_{t}$ and $p_{\phi}$ are constants of motion.  The motion of the test particle, neglecting the back reaction, is governed by the rest mass $m_0$, total energy $E$, axial angular momentum $L_z$ and Carter constant $\mathcal{Q}$, which is related to the second-rank irreducible tensor field of hidden symmetry \citep{Carter:1968rr}. To derive the geodesic equations in the first-order differential form, we employ the Hamilton-Jacobi equation following the integral approach pioneered by Carter \citep{Carter:1968rr}, which for the metric (\ref{metric}) read \citep{Chandrasekhar:1985kt}
\begin{align}
\Sigma \frac{dt}{d\lambda}=&\frac{r^2+a^2}{\Delta}(E(r^2+a^2)-aL_{z})-a(aE\sin^2{\theta}-L_{z}),\label{32}\\
\Sigma \frac{d\phi}{d\lambda}=&\frac{a}{\Delta}(E(r^2+a^2)-aL_{z})-(aE-\frac{L_z}{\sin^2{\theta}}),\label{33}\\
\Sigma \frac{dr}{d\lambda}=&\pm\sqrt{\mathcal{R}(r)}\ ,\label{req} \\
\Sigma \frac{d\theta}{d\lambda}=&\pm\sqrt{\Theta(\theta)}\ ,\label{theq}
\end{align}
where $\lambda$ is the affine parameter along the geodesics and the effective potentials $\mathcal{R} (r)$ and ${\Theta}(\theta)$ for radial and polar motion are given by 
\begin{align}
\mathcal{R}(r)=&E^2\left[\Big((r^2+a^2)-a\xi \Big)^2-\Delta \Big({\eta}+(a-{ \xi})^2\Big)\right],\label{Rpot}\\
\Theta(\theta)=&E^2[\eta-\left(\frac{{ \xi}^2}{\sin^2\theta}-a^2 \right)\cos^2\theta]\ . \label{theta0}
\end{align}

The separability constant $\mathcal{K}$ is related to the Carter constant $\mathcal{Q}$ by the equation $\mathcal{Q}=\mathcal{K}+(aE-L_z)^2$ \citep{Chandrasekhar:1985kt}. We introduce dimensionless quantities called impact parameters \citep{Chandrasekhar:1985kt}
\begin{eqnarray}
    \xi=\frac{L_z}{E} \;\; \text{,} \;\; \eta=\frac{\mathcal{K}}{E^2},
\end{eqnarray}
which are constant along the geodesics. The allowable zone for photon mobility around a black hole is  $\mathcal{R}\geq 0$ and $\Theta(\theta)\geq 0$ and the sign of $\dot{r}$ and $\dot{\theta}$ can be either positive or negative, chosen independently. At the turning points of motion, the sign change occurs, i.e. when $\mathcal{R}=0$ or $\Theta=0$ \citep{Chandrasekhar:1985kt}.
Our focus is on spherical photon orbits, which are spherical lightlike geodesics restricted on a sphere with a constant coordinate radius  $r$ characterized by $\dot{r}=0$ and $\ddot{r}=0$ \citep{Frolov:1418196,Chandrasekhar:1985kt}, which mathematically implies

\begin{eqnarray}
\mathcal{R}=\mathcal{R}'=0 \,\, \text{and}\,\,\mathcal{R}''\leq 0.\label{unstableOrbit} 
\end{eqnarray}

Solving Eq.~(\ref{unstableOrbit})  yields the critical values of impact parameters ($\xi_{c}, \eta_{c}$) for the unstable orbits, which read as
\begin{align}
\xi_{c}=& \frac{(a^2+r^2)\Delta '(r)-4r \Delta (r) }{a \Delta '(r)} \nonumber\\
\eta_{c}=&\frac{r^2 \left(8 \Delta (r) \left(2 a^2+r \Delta '(r)\right)-r^2 \Delta '(r)^2-16 \Delta (r)^2\right)}{a^2 \Delta '(r)^2}\label{CriImpPara},
\end{align}
where $'$ denotes the derivative concerning the radial coordinate.

 Photons with critical impact parameter $b = R_c$  are captured in an 
 unstable circular orbit, producing the lensed ``photon ring". In the Kerr metric \citep{Kerr:1963ud}, $R_c$ changes depending on the photon’s orientation relative to the spin axis, leading to a non-circular cross-section \cite{Bardeen:1973tla}. Although this variation is less than 4\%, it is potentially detectable \citep{Takahashi:2004xh, Johannsen:2010ru}. The unstable photon orbits have been investigated extensively for black holes and naked singularities \citep{Wilkins:1972rs,Goldstein1974,Johnston:1974pn,Izmailov1979,Izmailov1980,Teo:2020sey}, which are the boundaries between light ray capture and its scattered cross-section. Two circular photon orbits can exist in the equatorial plane of axially symmetric spacetimes: those that move in the black hole's rotational direction and those that move oppositely- referred to as prograde and retrograde photons, respectively. The Lens-Thirring effect \citep{Bardeen:1975zz} causes the rotation of a black hole to drag the inertial frame to an observer at infinity, resulting in shortened orbits for prograde photons to account for excess angular momentum. Conversely, the retrograde ones would need to revolve at larger radii since they had lost some angular momentum \citep{Bardeen:1972fi,Teo:2020sey}. Only when $\mathcal{\eta}_{c}>0$ do non-planar or three-dimensional photon orbits form; in contrast, when $\mathcal{\eta}_{c}=0$, photon orbits are planar and limited to the equatorial plane. The radii of prograde ($r_{p}^-$) and retrograde ($r_{p}^+$) orbit at the equatorial plane are obtained as roots of $\eta_{c}=0$, which for RQCBH  reads
\begin{eqnarray}\label{equiorbits}
  \left(r^4 -3 M r^3 + 3 \alpha  M^2\right)^2 -  4 a^2 M r^2 \left(r^3-2 \alpha  M\right) = 0,
\end{eqnarray}
which is difficult to solve analytically.
For $\alpha=0$, the solution of Eq.~(\ref{equiorbits}) reduces to the prograde ($r_{K}^-$) and retrograde ($r_{K}^+$) orbits of Kerr black holes 
which  are given as \citep{Teo:2020sey}
\begin{eqnarray}
    r_{K}^-\equiv2M\left[1+\cos\left({\frac{2}{3}\arccos\left(-\frac{|a|}{M}\right)}\right)\right]\ , \\
    r_{K}^+\equiv2M\left[1+\cos\left({\frac{2}{3}\arccos\left(\frac{|a|}{M}\right)}\right)\right]\ ,
\end{eqnarray}
and they fall in the range $M\leq r_K^-\leq 3M$ and  $3M\leq r_K^+\leq 4M$ \citep{Kumar:2018ple}. In the case of Schwarzschild black hole ($a=0$), the photon sphere of constant radius, $r_{p}^-=r_{p}^+=3M$ is formed when the two radii degenerate \citep{Kumar:2020hgm}. All spherical photon orbits are limited to the region $r_{p}^-<r_{p}<r_{p}^+$ when $r_{p}^+>r_{+}$. The Carter constant governs the additional latitudinal motion exhibited by the non-planar ($\theta\neq\pi/2$ and $\dot{\theta}\neq0$) geodesics in the $\eta_{c}>0$ situation, explaining a concealed spacetime symmetry. Additionally, these orbits cross the equatorial plane frequently while oscillating symmetrically about it. \citep{Teo:2020sey}. The values of   $r_{p}^-$ and $r_{p}^+$ for RQCBH, as tabulated in Table \ref{table1A}, show that the QC parameter  $\alpha$ introduces significant deviations from the predictions of GR. Specifically, $r_{p}^-$ and $r_{p}^+$ decrease as $\alpha$ increases, indicating that the quantum corrections shrink the photon orbits. For fixed spin values, the photon orbits for RQCBH are consistently smaller than those of the Kerr black hole, highlighting the impact of quantum corrections. 

\begingroup
\begin{table*}
	\caption{Prograde  ($r_{P}^{-}$) and retrograde  ($r_{P}^{+}$) photon orbit radii for RQCBH.}\label{table1A}
	\begin{ruledtabular}
		\begin{tabular}{l c c c c c c}
			\multicolumn{1}{c}{}&
            \multicolumn{2}{c}{Kerr Black Hole }&
			\multicolumn{2}{c}{ $\alpha=0.2M^2$}&
			\multicolumn{2}{c}{$\alpha= 0.6M^2$}\\
   \cmidrule{2-3} \cmidrule{4-5}  \cmidrule{6-7}
$a/M$&$r_{K}^{-}/M $ & $r_{K}^{+}/M $ & $r_{P}^{-}/M $ & $r_{P}^{+}/M $ &$r_{P}^{-}/M $ & $r_{P}^{+}/M $ \\
\hline \hline
 0. & 3. & 3. & 2.97726 & 2.97726 & 2.92832 & 2.92832 \\
 0.1 & 2.88219 & 3.11335 & 2.85696 & 3.09267 & 2.802 & 3.04856 \\
 0.2 & 2.75919 & 3.22281 & 2.73088 & 3.20386 & 2.66823 & 3.16374 \\
 0.3 & 2.63003 & 3.32885 & 2.59778 & 3.31138 & 2.52493 & 3.27459 \\
 0.4 & 2.49336 & 3.43184 & 2.45595 & 3.41563 & 2.36881 & 3.38169 \\
 0.5 & 2.3473 & 3.53209 & 2.30276 & 3.51698 & 2.19393 & 3.48548 \\
 0.6 & 2.18891 & 3.62985 & 2.13385 & 3.61571 & 1.98712 & 3.58634 \\
 0.7 & 2.01333 & 3.72535 & 1.94091 & 3.71206 & 1.7005 & 3.68456 \\
		\end{tabular}
	\end{ruledtabular}
\end{table*}
\endgroup

The black hole appears optically as the black hole shadow surrounded by the bright photon ring
\citep{Synge:1966okc,Bardeen:1973tla,Luminet:1979nyg,Cunningham:1973tf}.  The shadow has made it easier to estimate and measure various black hole properties, including mass, spin angular momentum and other hairs \citep{Kumar:2018ple}.  As a result, it can be used to test the no-hair theorem \citep{Cunha:2015yba} and GR in the strong-field regime  \citep{Gott:2018ocn,Kumar:2020owy}. The photon region surrounding the black hole's event horizon is formed by combining all unstable spherical photon orbits, i.e. the separatrix between photon geodesics that fall into the event horizon and those that escape to spatial infinity.  We suppose there are equally dispersed light sources at infinity, and photons arriving with all potential impact parameters are either caught by or scattered near the black hole. We additionally assume a distant observer at an inclination angle of $\theta_0$ with the black hole's rotating axis. The perceived angular distances of the image measured from the line of sight in directions parallel and perpendicular to the projected axis of rotation black hole onto the celestial sphere, respectively, are the celestial coordinates ($X$,$Y$) of the shadow boundary at the observer's sky \citep{Hioki:2009na}.

The relationship between two constants, $\xi_c$ and $\eta_c$ , and the image plane coordinates of the observer, $X$ and $Y$, is then determined. Using the tetrad components of the four-momentum $p^{(\mu)}$ and the geodesic Eqs.~(\ref{32}), (\ref{33}), and (\ref{theq}), for an observer at the position ($r_o,\theta_o$), we obtain:
\begin{eqnarray}
&&X= -r_o\frac{p^{(\phi)}}{p^{(t)}} = -\left. r_o\frac{\xi_c}{\sqrt{g_{\phi\phi}}(\zeta-\gamma\xi_c)}\right|_{(r_o,\theta_o)},\nonumber\\
&&Y = r_o\frac{p^{(\theta)}}{p^{(t)}} =\pm\left. r_o\frac{\sqrt{\mathcal{V}_{\theta}(\theta)}
}{\sqrt{g_{\theta\theta}}(\zeta-\gamma\xi_c)}\right|_{(r_o,\theta_o)},~\label{Celestial}
\end{eqnarray} 
where 
\begin{eqnarray}
\zeta=\sqrt{\frac{g_{\phi\phi}}{g_{t\phi}^2-g_{tt}g_{\phi\phi}}},\qquad \gamma=-\frac{g_{t\phi}}{g_{\phi\phi}}\zeta.
\end{eqnarray}
The coordinates $X$ and $Y$ in Eq.~(\ref{Celestial}) represent the apparent displacement along the parallel and perpendicular  axes to the projected axis of the black hole symmetry. The celestial coordinate, Eq.~(\ref{Celestial}) for an observer in an asymptotically flat region ($r_o\to\infty$), yields \citep{bardeen1973}.

\begin{equation}
X=-\xi_c\csc\theta_o,\qquad Y=\pm\sqrt{\eta_c+a^2\cos^2\theta_o-\xi_c^2\cot^2\theta_o}.\label{pt}
\end{equation} 
Further, if the observer is at the equatorial plane ($\theta_o=\pi/2$), it simplifies to
\begin{eqnarray}
X=-\xi_{c}\ , \nonumber\\ 
Y=\pm\sqrt{\eta_{c}}\ .\label{Celestial3}
\end{eqnarray}

\begin{figure*}[hbt!]
\centering
\begin{tabular}{p{7.5cm} p{7.5cm}}
  
    \includegraphics[scale=0.855]{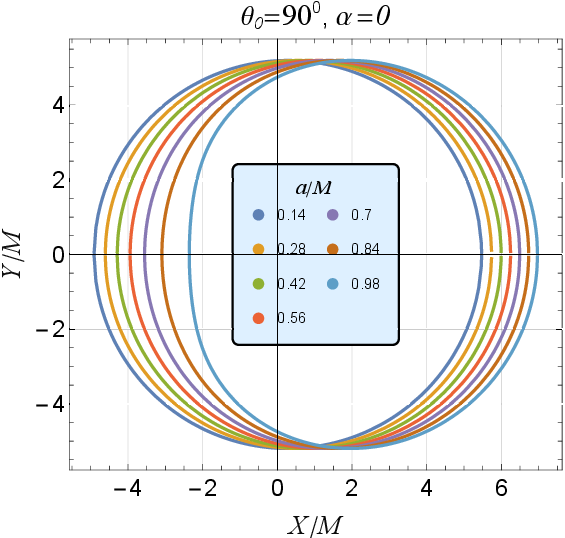}
    \end{tabular}
  \caption{Shadow silhouettes of Kerr black hole ($\alpha=0$) with varying spin $a/M$ at an inclination angle  $90$\textdegree (right).}\label{shadow1}	
 \end{figure*}

\begin{figure*}[hbt!]
\centering
\begin{tabular}{p{9cm} p{9cm}}
 
    \includegraphics[scale=0.9]{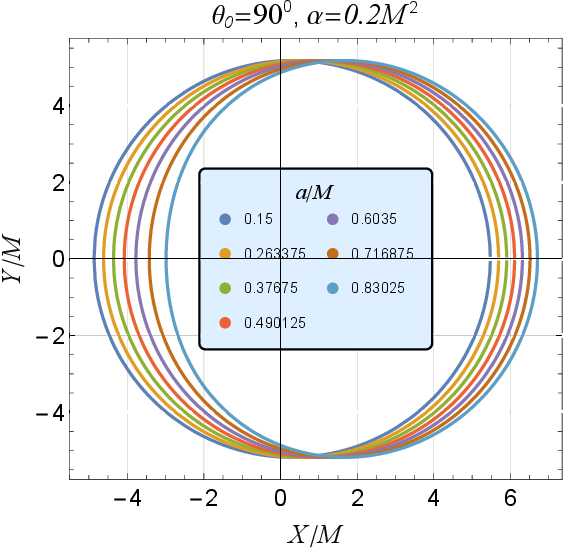}&
    \includegraphics[scale=0.9]{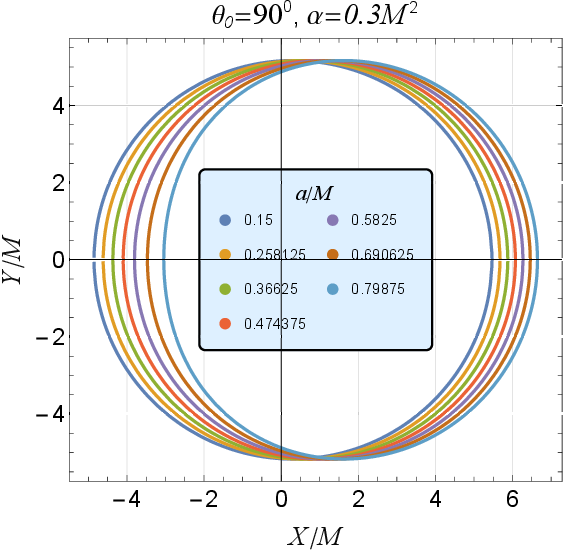}\\
    \end{tabular}

  \caption{Shadow silhouettes of the RQCBH  with varying parameters $a$ and $\alpha$ for an inclination angle $\theta_0 = 90$\textdegree~.}\label{shadow}	
 \end{figure*}

The Eq.~(\ref{Celestial3}) ensures that the shadow of a non-rotating black hole (\ref{metric1}) is perfectly circular, and further, Eq.~(\ref{Celestial3}) reverts to $X^2+ Y^2=27M^2$ for the Schwarzschild black hole. The shadows for the Kerr black holes ($\alpha=0$) are illustrated in Fig.~\ref{shadow1}, and we also compare the shadows of the RQCBH with those of the Kerr black hole (cf. Fig.\ref{shadow}). The parameter $\alpha$ is observed to have a decreasing effect on the size of the RQCBH shadow, as for fixed values of $a$, the shadow size diminishes with increasing $\alpha$.

A mathematically compatible rotating black hole model in QG is lacking. The lack of rotating black hole solutions in QG substantially hampers the development of testing QG from observations, e.g., the EHT observations. The EHT observation revealed event horizon-scale images of the supermassive black holes Sgr A* and M87*. We present RQCBH spacetimes with an additional QC parameter $\alpha$ and constrain it by EHT observations. Recent studies have examined quantum corrections to rotating black holes, including QG motivated versions of the Kerr black hole \citep{Liu:2020ola,Brahma:2020eos,KumarWalia:2022ddq,Islam:2022wck,Afrin:2022ztr, Yang:2022btw}. These corrections significantly impact observable properties like shadow shape and size, light deflection, and signal time delays.  
We found that the shadow silhouette is constructed by plotting ($X$, $Y$) parametrically as a function of $r_p$. Figs.~\ref{shadow1} and \ref{shadow} reveal that the shadow shrinks and becomes increasingly distorted as the quantum parameter $\alpha$ diminishes. It demonstrates that the QC parameter, though expected to be relevant only at the Planck scale, has a significant and measurable impact on observable features such as the shadow's shape and size. 
We exploit this visible effect to see whether these imprints of the $\alpha$ in the shadow can be exploited to extract and constrain the parameter $\alpha$. Additionally, there is a horizontal shift in the shadow centre along the $X$-axis, with an increasing $a$, because of the frame-dragging effect.

 \section{Observables and black hole parameter estimation}\label{Sec5}
\begin{figure*}[hbt!]
		\begin{tabular}{c c}
			\includegraphics[scale=0.9]{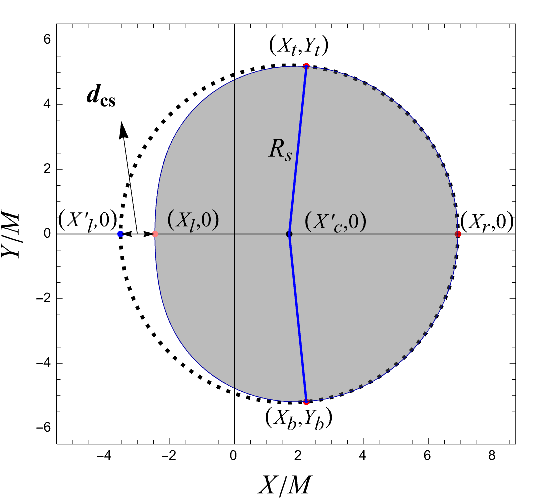}
        \end{tabular}
       \caption{The shadow observables, radius $R_s$ and distortion $\delta_s = d_{cs}/R_s$, for the apparent shape of the shadow of the RQCBH ($a = 0.95M$, $\alpha = 0.061$, and $\theta_0=\pi/2$). The distortion is the difference between the left endpoints of the dotted circle and of the shadow. }
	\label{circle}
\end{figure*}

The parameters associated with the RQCBHs are expected to be constrained from EHT observations. The supermassive black holes M87* and Sgr A*'s observed images match the GR's Kerr black hole. Nevertheless, black holes in MTG were not mentioned in the EHT observation \citep{EventHorizonTelescope:2019dse,EventHorizonTelescope:2019ths,EventHorizonTelescope:2019ggy}, which is what we plan to investigate using the calculated parameters of our RQCBHs. We assume the observer is in the equatorial plane  ($\theta_0=\pi/2$) for parameter estimation.
\paragraph{Hioki and Maeda method}:
Hioki and Maeda (\citeyear{Hioki:2009na}) proposed the two observables, $R_s$ and $\delta_s$, to quantify the size and distortion of the black hole shadow. A reference circle approximates the shadow, and its radius is denoted by $R_s$. The left edge of the shadow's divergence from the circle boundary is represented by $\delta_s$ \citep{Hioki:2009na}. The shadow reference circle and the shadow contour coincide at the top ($X_t$, $Y_t$), bottom ($X_b$, $Y_b$)), and right ($X_r, 0$)) edges, respectively \citep{Ghosh:2020ece}. The positions where the reference circle and the leftmost edge of the shadow contour meet the horizontal axis are ($X^{'}_l$, 0) and ($X_l$, 0), respectively (cf. Fig.~\ref{circle}). Indeed, $\delta_s$ indicates the shadow's deviation from the reference circle, whereas $R_s$ indicates the estimated size of the shadow. 
These observables are defined as \citep{Hioki:2009na}
\begin{eqnarray}
    R_s=\frac{(X_t-X_r)^2+Y_{t}^2}{2|X_r-X_t|},
\end{eqnarray}\label{Rs}
using the relations $X_b=X_t$ and $Y_b=-Y_t$, and
\begin{eqnarray}
    \delta_s=\frac{|X_l-X^{'}_l|}{R_s}\ ,
\end{eqnarray}\label{deltas}
where subscripts $r$, $l$, $t$ and $b$, respectively, stand for the right, left, top and bottom of the shadow boundary.

\begin{figure*}[hbt!]
		\begin{tabular}{c c}
			\includegraphics[scale=0.72]{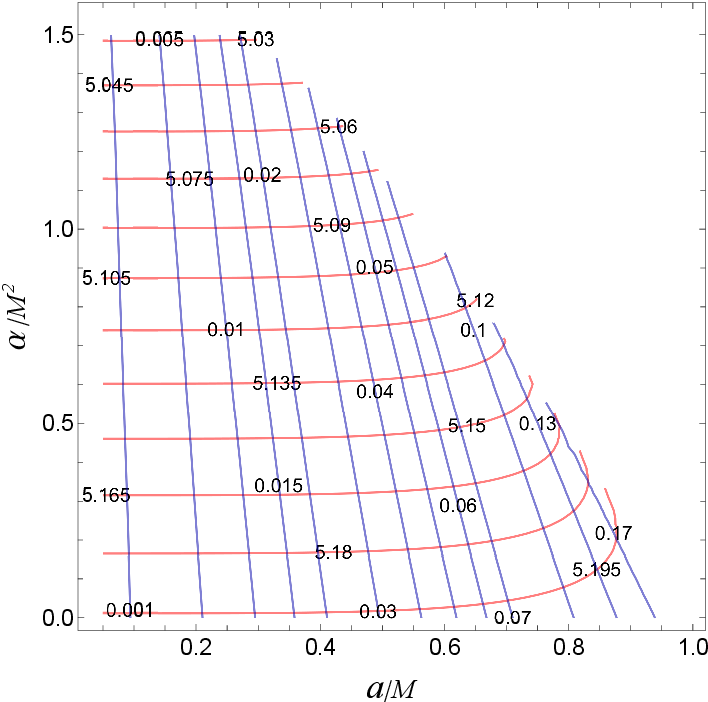}
        \end{tabular}
	\caption{Contour plots of the shadow observables  $Rs$ (red curves) and $\delta$s (blue curves) in the RQCBH parameter space ($a/M$, $\alpha/M^2$). The contour intersection points uniquely determine RQCBH parameters. }
	\label{parameterestimation1}
\end{figure*}

Plotting the contours of $R_s$ and $\delta_s$ in the $(\alpha-a)$ parameter space of the RQCBH in Fig.~\ref{parameterestimation1} allows us to estimate the black hole parameters. The parameters $a$ and $\alpha$ are uniquely determined by an intersection point of $R_s$ and $\delta_s$ (cf.~\ref{parameter_table1}).

\begin{table*}[hbt!]
\begin{tabular}{|p{1.6cm}|p{1.6cm}|p{1.6cm}|p{1.6cm}| }
    \hline 
$R_s$ & $\delta$s & $a$/M & $\alpha/M^2 $\\
\hline \hline
5.03 & 0.001 & 0.0642 & 1.485 \\ 
\hline
5.06 & 0.03 & 0.3518 & 1.258 \\
\hline
5.105 & 0.06 & 0.5226 & 0.8943 \\
\hline
5.15 & 0.01 & 0.2669 & 0.4622 \\ 
\hline
5.15 & 0.1 & 0.7096 & 0.561 \\
\hline
5.195 & 0.005 & 0.2094 & 0.01323 \\
\hline
5.195 & 0.02 & 0.3587 & 0.01572\\
\hline
\end{tabular}
\caption{Estimated values of RQCBH parameters $\alpha/M^2$ and $a/M$ from known shadow observables $R_s$ and $\delta$s at inclination angle $\theta=90$\textdegree~.}\label{parameter_table1}
\end{table*}

\paragraph{Kumar and Ghosh method:}
The Hioki and Maeda method \citep{Hioki:2009na} of numerical estimation of black hole parameters using shadow observables $R_s$ and $\delta_s$ was later extended by \cite{Tsupko:2017rdo} to analytical estimation. Additionally, \cite{Tsukamoto:2014tja} put out a strategy for differentiating the spinning black hole shadows emerging in the MTG from the Kerr black hole shadow. Nevertheless, the two approaches mentioned above need certain symmetries in the shadow shape for the observables $R_s$ and  $\delta_s$, which might not be accurate for the haphazard black hole shadows in MTG \citep{Schee:2008kz,Johannsen:2015qca,Tsukamoto:2014tja, Abdujabbarov:2015xqa,Younsi:2016azx, Tsupko:2017rdo}. Additionally, noisy data may cause the shadow shape to be deviated from exactly circular \citep{Abdujabbarov:2015xqa,Kumar:2018ple}. This served as motivation for \cite{Kumar:2018ple}, who used observables describing a haphazard (or non-circular) shadow to estimate the parameters related to black holes. They defined the shadow Area ($A$) and oblateness ($D$) by
\begin{eqnarray}
A=2\int{Y(r_p) dX(r_p)}=2\int_{r_p^{-}}^{r_p^+}\left( Y(r_p) \frac{dX(r_p)}{dr_p}\right)dr_p,\label{Area}
\end{eqnarray} 
and
\begin{eqnarray}
D=\frac{X_r-X_l}{Y_t-Y_b}.\label{Oblateness}
\end{eqnarray}
For an equatorial observer, the oblateness $D$ may vary between $\sqrt{3}/2\leq D<1$, with $D=1$ for the Schwarzschild case ($a=0$) and $D=\sqrt{3}/2$ for extremal Kerr case ($a=M$) \citep{Tsupko:2017rdo}. 

\begin{figure*}
		\begin{tabular}{c c}
			\includegraphics[scale=0.72]{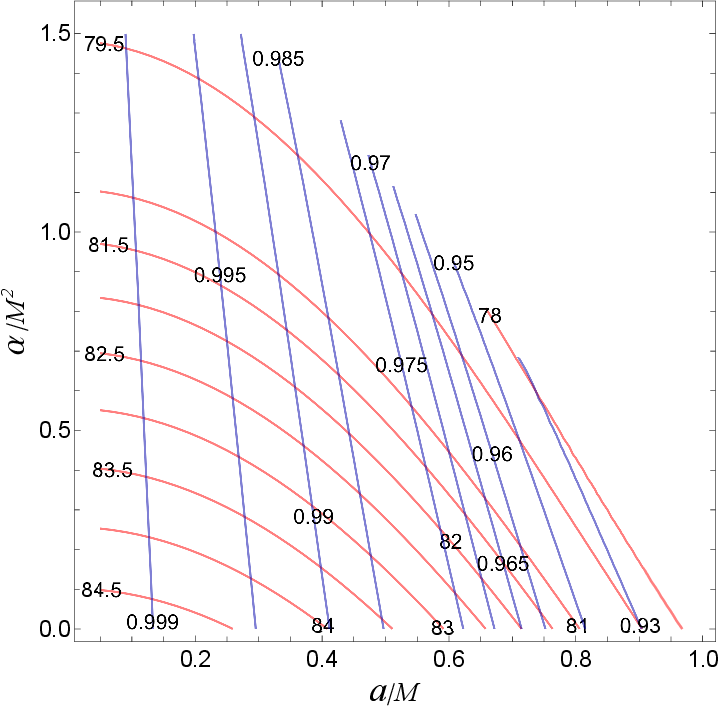}
		\end{tabular}
	\caption{Contour plots of the shadow observables  $A$  and $\delta$ in the RQCBH parameter space ($a/M$, $\alpha/M^2$) .  Red curves correspond to constant $A$, and blue curves are for constant oblateness parameter $D$}
	\label{parameterestimation2}
\end{figure*}

Furthermore, we plot the contour map of the observables $A$ and $D$ in the ($\alpha$-$a$) plane (cf. Fig.~\ref{parameterestimation2}). Fig.~\ref{parameterestimation2} shows that the intersection points provide exact values for parameters a and $\alpha$ based on A and D values. Table~\ref{parameter_table2} displays the estimated values for RQCBH parameters $a$ and $\alpha$ based on shadow observables $A$ and $D$.

\begin{table*}
 
\begin{tabular}{|p{1.6cm}|p{1.6cm}|p{1.6cm}|p{1.6cm}| }
\hline
A & D & $a$/M & $\alpha/M^2$ \\
\hline\hline 
81 & 0.96 & 0.7115 & 0.2099 \\ 
\hline 
81 & 0.97 & 0.5998 & 0.4475 \\
\hline 
82 & 0.97 & 0.6489 & 0.1291 \\
\hline 
82 & 0.995 & 0.2511 & 0.7268 \\
\hline 
83 & 0.985 & 0.4762 & 0.1874 \\
\hline 
83 & 0.995 & 0.2684 & 0.4305 \\
\hline 
84 & 0.995 & 0.2859 & 0.1311\\
\hline
\end{tabular}
\caption{Estimated values of RQCBH parameters $\alpha/M^2$ and $a/M$ from known shadow observables $A$ and $D$ at inclination angle $\theta=90$\textdegree~.}\label{parameter_table2}
\end{table*} 
	
\section{Constraints from the \textit{EHT} observations of supermassive black holes of M87* and Sgr A*}\label{Sec6}
The black hole shadow boundary forms from photons escaping the gravitational field, showcasing strong-field aspects \citep{Jaroszynski:1997bw,Falcke:1999pj}.

The EHT collaboration has released images of SMBHs M87* \citep{EventHorizonTelescope:2019dse,EventHorizonTelescope:2019ggy} and Sgr A* \citep{EventHorizonTelescope:2022wkp,EventHorizonTelescope:2022xqj}, which are largely consistent with the shadows predicted by Kerr black holes, supporting the Kerr hypothesis \citep{Psaltis:2007cw}. However, due to the uncertainties in the EHT results, MTG theories cannot be entirely ruled out. It opens the possibility of testing the no-hair theorem \citep{Carter:1971zc} and exploring quantum correction in the strong-field regime. Shadow characteristics, described by various observables, reveal details of the underlying spacetime. We utilize the EHT bounds on the Schwarzschild deviation ($\delta$) of the shadow for M87* and Sgr A*, modelling them as RQCBHs to evaluate the viability of quantum gravity at the current EHT resolution.
The Schwarzschild deviation is computed by first calculating the shadow area, given by \citep{Kumar:2018ple,Afrin:2021imp}: 
\begin{equation}\label{Area1} 
A=2\int_{r_p^{-}}^{r_p^+}\left( Y(r_p) \frac{dX(r_p)}{dr_p}\right)dr_p.
\end{equation} 
The deviation ($\delta$) between the model shadow diameter ($\Tilde{d}{metric}$) and the Schwarzschild diameter $6\sqrt{3}M$ is \citep{EventHorizonTelescope:2022wkp,EventHorizonTelescope:2022xqj}: 
\begin{equation}\label{SchwarzschildShadowDiameter} \delta=\frac{\Tilde{d}_{metric}}{6\sqrt{3}}-1,\;\; \Tilde{d}_{metric}=2R_a. \end{equation}
Then the angular shadow diameter, for a distance $d$ from the black hole, is defined as
\citep{Bambi:2019tjh,Kumar:2020owy,Afrin:2021imp}
\begin{eqnarray}
d_{sh}=2\frac{R_a}{d}\;,\;R_a=\sqrt{A/\pi},\label{angularDiameterEq}
\end{eqnarray}  
$R_a$ being the areal shadow radius. Apart from $d$, $d_{sh}$ also depends on the mass $M$ of the black hole. As $a/M$ varies from 0 to 1 and $\theta_0$ from 0 to $90^\circ$, Kerr black hole shadows can shrink up to 7.5\% compared to the Schwarzschild shadow diameter, with $-0.075\leq\delta\leq0$ indicating consistency with Kerr. At the same time, deviations outside this range suggest otherwise \citep{EventHorizonTelescope:2022xqj}. RQCBHs exhibit smaller and larger shadows than Kerr shadows for specific parameters ($\alpha, a$), making them testable SMBH candidates with $\delta$ bounds.

While the EHT observations provide extensive details about the image of M87* \citep{EventHorizonTelescope:2019dse} and Sgr A* \citep{EventHorizonTelescope:2022wkp}, to constrain the QC parameter, we will focus solely on two key observables: the angular shadow diameter $d_{sh}$ and the Schwarzschild shadow deviation $\delta$.

\subsection{Constraints from the \textit{EHT} results of M87*}
The supermassive black hole M87*'s event horizon scale image was made public by the \textit{EHT} project utilizing VLBI technology \citep{EventHorizonTelescope:2019dse,EventHorizonTelescope:2019ths,EventHorizonTelescope:2019ggy}.  The observed image of M87* is crescent-shaped and almost circular; the angular dimension $\theta_d$ of the emission zone in the observed image is $42 \pm 3\mu as$. The error $\pm 3\mu as$ represents the measurement uncertainty from observation. The simulated appearance of the Kerr black hole constrains the observed image \citep{EventHorizonTelescope:2019dse,EventHorizonTelescope:2019ths,EventHorizonTelescope:2019ggy}. It can also be applied to constrain the MoG black hole models. 

 \begin{figure*}[hbt!]

\begin{tabular}{c c}
\includegraphics[scale=0.72]{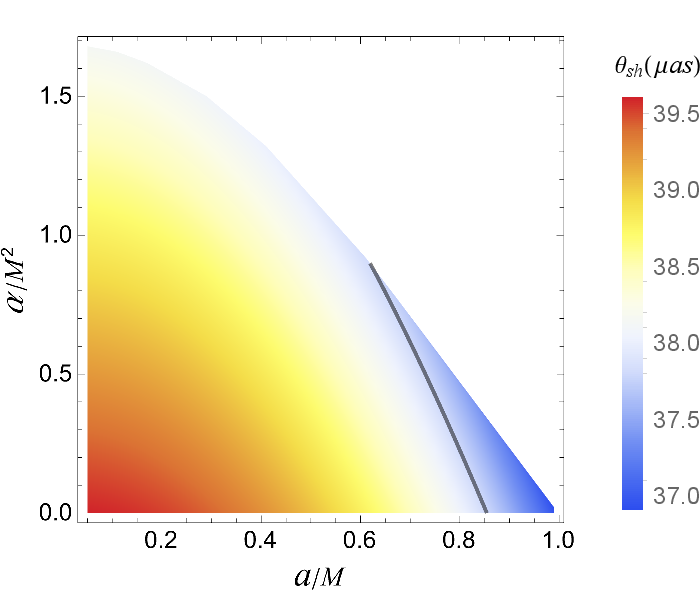}
\includegraphics[scale=0.72]{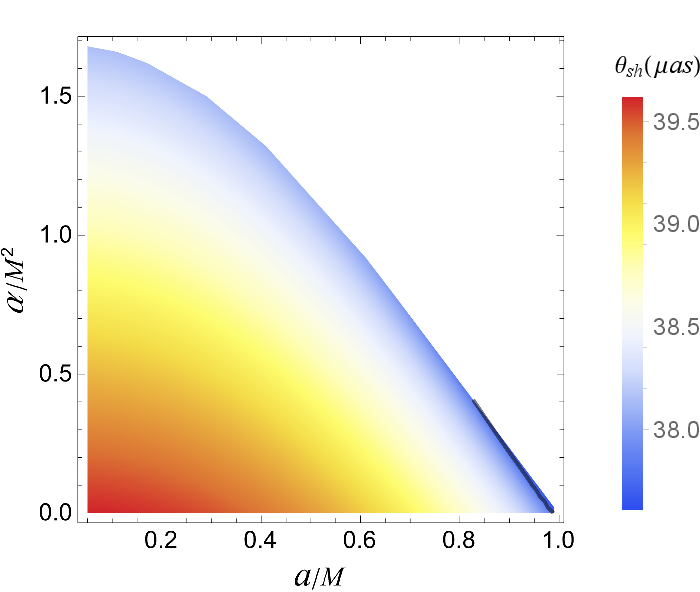}
		\end{tabular}	
  \caption{Angular diameter observable $\theta_{sh}$, in units of $\mu$as, for RQCBH shadows as a function of parameters ($a/M$, $\alpha/M^2$) with black solid curve corresponding to $\theta_{sh}=37.8\mu$as of the M87* black hole. The substantial region is consistent with the EHT observation of M87*, and the white region is forbidden for ($a/M$, $\alpha/M^2$). The inclination angle is $\theta_0=17$\textdegree~$(left)$ and $\theta_0=90$\textdegree~$(right)$.}
	\label{M87obs2}
	
 \end{figure*}
Various observational systematics can obscure EHT data due to the sparse telescope array and uncertainties in the underlying astrophysics of radiation, plasma, and accretion phenomena \citep{Gralla:2020pra,Gralla:2019xty}. Additionally, it remains unclear if the ring-like features in M87* and Sgr A* images are gravitationally lensed photon regions, the surrounding accretion disk, or both \citep{Gralla:2020pra,Gralla:2019xty}. We treat the bright ring enclosing the dark region as the shadow and calculate its observables.
When considering the position of M87*'s relativistic jet, the EHT estimated the inclination angle to the line of sight at approximately $163$\textdegree~ \citep{Walker:2018muw}. Since our analysis uses an analytical shadow curve without accretion flow effects, the shadow's top-bottom symmetry implies that an inclination of $163$\textdegree~ is roughly equivalent to $17$\textdegree. However, maximum shadow distortion occurs at high inclinations ($\theta_0\approx90$\textdegree), so we conduct our analysis at observation angles of $\theta_0=0$\textdegree~,  17\textdegree~ and 90\textdegree.
\begin{figure*}[hbt!]
\begin{center}
\begin{tabular}{c c}
\includegraphics[scale=0.72]{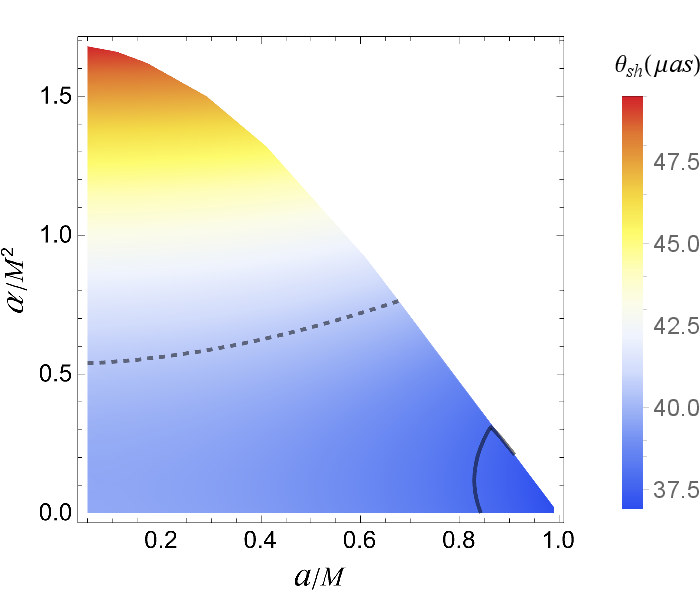}
\includegraphics[scale=0.72]{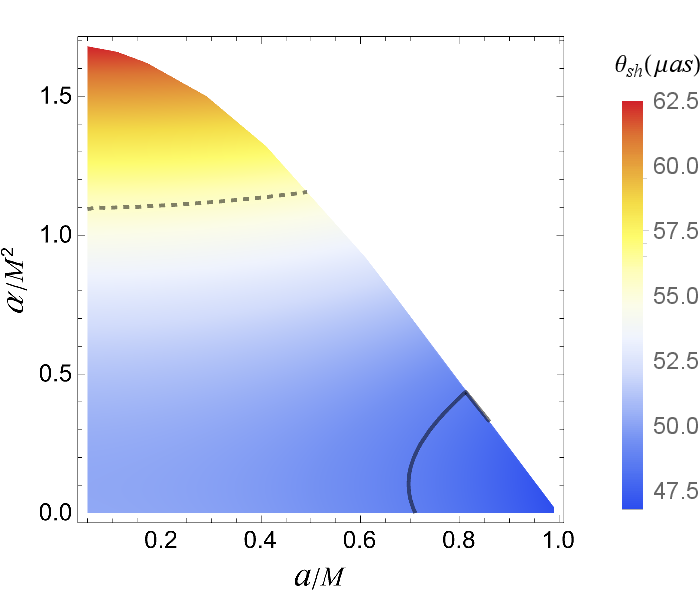}
\end{tabular}	
\end{center}
\caption{Angular diameter observable $\theta_{sh}$ for RQCBH shadows as a function of parameters ($a/M$, $\alpha/M^2$). The solid and the dashed curves correspond respectively to $\theta_{sh}=37.8\mu as$ and $40.5\mu as$ of the M87* black hole (left) and $\theta_{sh}=48.7\mu as$, $55.7\mu as$ of the SgrA* black hole (right). The substantial regions cast shadows consistent with the M87* and SgrA* shadow sizes of $EHT$, and the white region is forbidden for ($a/M$, $\alpha/M^2$).  The inclination angle is $\theta_0 \approx$ 0\textdegree~.}
	\label{zerodegree}
\end{figure*}
\begin{figure}[hbt!]
\begin{center}
\includegraphics[scale=0.75]{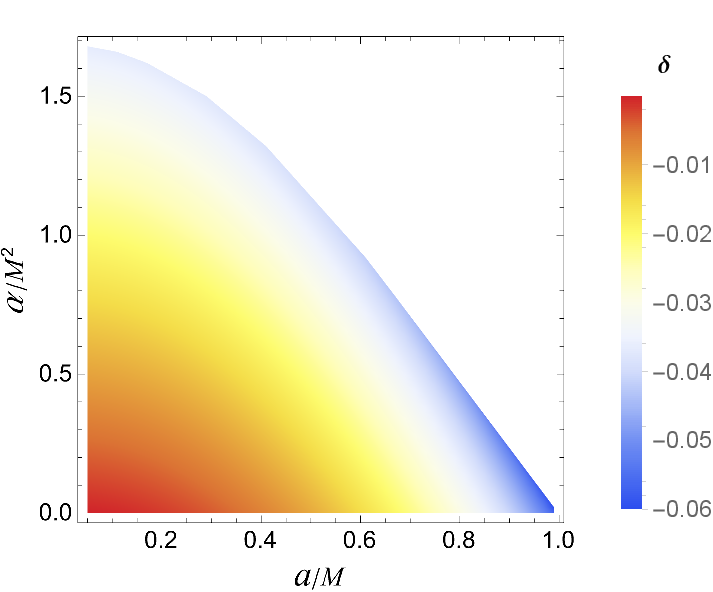}
\caption{ Plot showing Schwarzschild deviation $\delta$ of the RQCBH shadows. It agrees with the EHT observations of the M87* black hole $\delta_{M87^*}=-0.01\pm0.17$. The inclination angle is $\theta_0=17$\textdegree~. The white region pertains to no-horizon spacetime. }\label{m872}
\end{center}	
\end{figure}

Considering $\leq 10 \%$  offset between the diameter of the ring and photon ring, the angular diameter of the shadow of M87* can be calibrated to $\theta_{sh}=37.8\pm2.7\mu as$ \citep{EventHorizonTelescope:2019dse, Banerjee:2022bxg}. The error $\pm2.7\%$ incorporates both measurement uncertainty and potential offset. In Fig.~\ref{M87obs2}, we depict the calculation of RQCBH shadow angular diameter of M87* for $\theta_o=17$\textdegree~ and 90\textdegree~  as a function of ($a/M$, $\alpha/M^2$). The entire parameter space aligns with the EHT observations within the $1\sigma$ range, $[35.1-40.5]\mu$as. However, to establish bounds on ($a/M$, $\alpha/M^2$), we adopt a conservative approach by considering the median value of $\theta_{sh}$ up to the maximum admitted by the RQCBH model. Specifically, we select $\theta_{sh} \geq 37.8 \mu$as. The M87* black hole shadow angular diameter places a bound on the parameters ($a$, $\alpha$) where $a \in (0, 0.8511 M)$ at $\alpha=0$  and $a \in (0, 0.6157 M)$ at $\alpha=0.8985 M^2$, for $\theta_o=17$\textdegree~ and $a \in (0, 0.8262 M)$ at $\alpha=0$  and $a \in (0, 0.9799 M)$ at $\alpha=0.4141 M^2$ for $\theta_o=90$\textdegree~ (cf. Fig.~\ref{M87obs2}). The shadow angular diameter for RQCBH is also calculated for $\theta_o\approx0$\textdegree~ as a function of ($a/M$, $\alpha/M^2$) and depicted in  Fig. \ref{zerodegree}. Imposing bounds on $\theta_{sh}$, the limits come out to be $a \in (0, 0.8427 M]$ at $\alpha=0.00939 M^2$ and $a \in (0, 0.8684 M]$ at $\alpha=0.3128 M^2$ for $\theta_{sh}\geq37.8 \mu as$, $a \in (0, 0.04592 M]$ at $\alpha=0.5423 M^2$ and $a \in (0, 0.6733 M]$ at $\alpha=0.7656 M^2$ for $\theta_{sh}\leq40.5 \mu as$ (cf. Fig.~\ref{zerodegree}).

The first EHT image of M87* showed an asymmetric bright ring, formed by strong lensing and relativistic beaming, and a central brightness depression identified as the black hole shadow \citep{EventHorizonTelescope:2019dse}. The M87* ring's diameter, calibrated with the shadow size provides a Schwarzschild deviation of $\delta_{M87^*}=-0.01\pm0.17$ at $1\sigma$ confidence \citep{EventHorizonTelescope:2021dqv,Afrin:2024khy}. We use the estimated mass of the black hole $M_{M87^*} = 6.5\times10^9 M_\odot$ and its distance from Earth $d_{M87^*} = 16.8 Mpc$ to impose $\delta_{M87^*}$ with the EHT inferred bound. The whole parameter space in the case of the RQCBH is satisfied when modelling M87* as an RQCBH when the EHT inferred bound is imposed (cf. Fig.~\ref{m872}).

 Thus, the M87* can be RQCBH spacetime in this constrained parameter space. Since several possible parameter points ($a$, $\alpha$) exist within the confined parameter space, the RQCBH's compatibility with the M87* observations demonstrates that they can be excellent candidates for astrophysical black holes \citep{EventHorizonTelescope:2019ggy,EventHorizonTelescope:2021dqv}.

\subsection{Observational constraints from the EHT results of Sgr A*}
The shadow data for the Sgr A* black hole was provided by the EHT collaboration \citep{EventHorizonTelescope:2022exc,EventHorizonTelescope:2022urf,EventHorizonTelescope:2022apq,EventHorizonTelescope:2022wok,EventHorizonTelescope:2022wkp, EventHorizonTelescope:2022xqj} based on the 2017 VLBI observing mission at 1.3 mm wavelength. Images of the shadow of Sgr A* black hole are useful for determining the characteristics of astrophysical black holes as  (i) Sgr~A* probes a $10^{6}$ order of higher curvature than the M87* and (ii)  separate previous estimations for the mass-to-distance ratio are used. The shadow images were created using various imaging and modelling techniques, and they are astonishingly similar in features.

The mass and distance of Sgr A* have been determined as $M = 4.0_{-0.6}^{+1.1} \times 10^6 M\odot$ and $d = 8$ kpc \citep{EventHorizonTelescope:2022wkp,EventHorizonTelescope:2022xqj}, based on independent stellar dynamic observations of the S0-2 star's orbit using the Keck telescopes and the Very Large Telescope Interferometer (VLTI) \citep{Do:2019txf,Chael:2021rjo,Nicolas:2023cnc,EventHorizonTelescope:2022xqj}. The EHT image of Sgr A* shows an angular shadow diameter of $d_{sh} = 48.7 \pm 7,\mu$as, with a Schwarzschild deviation of $\delta_{Sgr A^*} = -0.08_{-0.09}^{+0.09}$ (VLTI) and $-0.04_{-0.10}^{+0.09}$ (Keck), consistent with the expected shadow of a Kerr black hole \citep{EventHorizonTelescope:2022wkp,EventHorizonTelescope:2022xqj}. While the inclination angle remains uncertain, an inclination greater than $50$\textdegree, has been disfavored \citep{EventHorizonTelescope:2022xqj}.

\begin{figure*}[hbt!]
\begin{center}
\begin{tabular}{c c}
\includegraphics[scale=0.72]{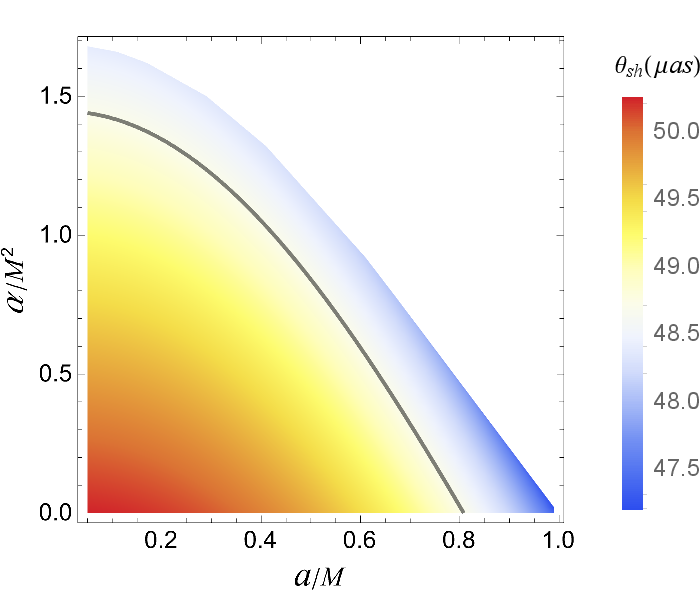}
\includegraphics[scale=0.72]{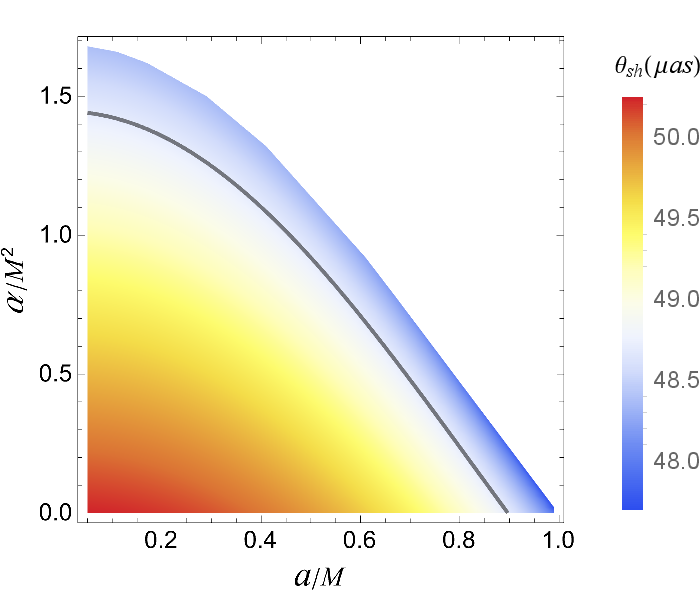}
\end{tabular}	
\end{center}

\caption{Angular diameter observable $\theta_{sh}$, in units of $\mu as$, for RQCBH shadows as a function of parameters ($a/M$, $\alpha/M^2$). The black solid curve corresponds to $\theta_{sh}=48.7\mu as$ of the SgrA* black hole reported by the $EHT$. The substantial region casts a shadow that is consistent with the SgrA* shadow size of $EHT$ and the white region is forbidden for ($a/M$, $\alpha/M^2$).  The inclination angle is $\theta_0=$50\textdegree~(left) and $\theta_0=$90\textdegree~(right).}
	\label{M87obs3}
\end{figure*}

We determine $\theta_{sh}$, which, in addition to other black hole parameters, relies on the mass $M$, the distance $d$ of the black hole, QC parameter $\alpha$, and inclination $\theta_0$ as depicted in Fig.~\ref{M87obs3}. The EHT collaboration using the calibration method has obtained a shadow angular diameter of $SgrA^*= 48.7 {\pm} 7\mu as$.  We shall use mean shadow diameter, $d_{sh}^{Sgr A*}$ as 48.7 $\mu as$.

 In Fig.~\ref{M87obs3}, we calculate the RQCBH shadow angular diameter of SgrA* for $\theta_o=50$\textdegree~ and 90\textdegree~  as a function of ($a/M$, $\alpha/M^2$). The shadow angular diameter $\theta_{sh}$ as a function of ($a/M$, $\alpha/M^2$) is in agreement with EHT observation of SgrA*(cf. Fig  \ref{M87obs3}. The $\theta_{sh} \in [41.7\mu as, 55.7 \mu as ]$  is satisfied for the entire parameter space ($a/M$, $\alpha/M^2$).  We use conservative approach by considering the median value of $\theta_{sh}$ up to the maximum admitted by the RQCBH model to put bounds and take $\theta_{sh}\geq48.7\mu as$. This range of angular diameter strongly constrains the parameters  ($a$, $\alpha$) such that for $0.0 \leq \alpha\leq 1.443 M^2$, the allowed range of $a$ is (0,0.8066 M) for $\theta_o=50$\textdegree~ and for $0.0 \leq \alpha\leq1.447 M^2$, the allowed range of $a$ is $(0,0.894M)$  for $\theta_o=90$\textdegree~ (cf. Fig  \ref{M87obs3}). Similarly, the shadow angular diameter for RQCBH is also calculated for $\theta_o\approx0$\textdegree~ as a function of ($a/M$, $\alpha/M^2$) and depicted in  Fig.~\ref{zerodegree}. Imposing the already discussed bounds on $\theta_{sh}$, the limits come out to be $a \in (0, 0.7082 M]$ at $\alpha=0.005488 M^2$ and $a \in (0, 0.8177 M]$ at $\alpha=0.4417 M^2$ for $\theta_{sh}\geq48.7 \mu as$, $a \in (0, 0.04797 M]$ at $\alpha=1.106 M^2$ and $a \in (0, 0.48462 M]$ at $\alpha=1.164 M^2$ for $\theta_{sh}\leq55.7 \mu as$ (cf. Fig.~\ref{zerodegree}). For this constrained parameter range, the RQCBH shadows are consistent with the shadow of Sgr A*.
 
For Kerr black hole with $a\leq M$, the Schwarzschild shadow deviation lies at $-0.075\leq  \delta \leq 0$ as the inclination varies from 0 to $\pi/2$.
EHT used the two separate priors for the Sgr~A* angular size from the Keck and Very Large Telescope Interferometer (VLTI) observations to estimate the bounds on the fraction deviation observable $\delta$  \citep{EventHorizonTelescope:2022wkp,EventHorizonTelescope:2022urf}

\begin{align}
\delta_{Sgr}= \begin{dcases*} -0.08^{+0.09}_{-0.09}\;\;\;\;\; & \text{VLTI}\\
-0.04^{+0.09}_{-0.10}\;\;\;\;\; &\text{Keck}	 
\end{dcases*} 
\end{align}

\begin{figure}[hbt!]
\begin{center}
	\includegraphics[scale=0.75]{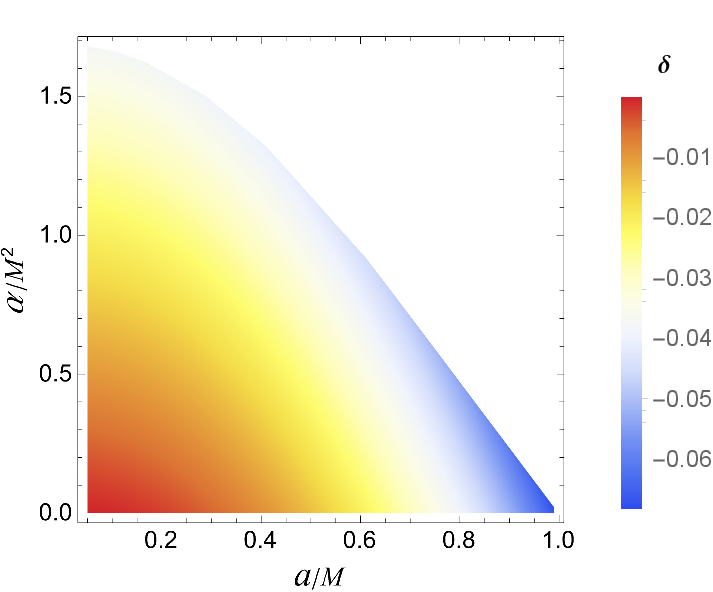}
	\caption{ Plot showing Schwarzschild deviation $\delta$ of the RQCBH shadows. It is in agreement with the EHT observations of the Sgr A* black hole $\delta_{SgrA^*}=-0.08_{-0.09}^{+0.09}$ (VLTI), $-0.04_{-0.10}^{+0.09}$ (Keck). The inclination angle is $\theta_0=50$\textdegree~, and the white region is forbidden for ($a/M$, $\alpha/M^2$). }\label{sgr2}

\end{center}	
\end{figure}
We set the bounds on the RQCBH parameters and impose the inferred limitations of EHT  on $\delta$. We report that modelling Sgr A* as an RQCBH satisfies the entire parameter space for inferred bound Keck (-0.14,0.05) and VLTI  (-0.17,0.01) observations. Strong evidence that the RQCBHs can be viable candidates for another supermassive black hole is provided by the parameter space's good consistency with the observational results for Sgr A* (cf. Fig.~\ref{sgr2}).

\begin{table}[h!]
\centering
\begin{tabular}{|c|c|c|}
\hline
\multirow{2}{*}{\textbf{Rotating black hole metric in modified theories of gravity}} &  \multirow{2}{*}{\textbf{Constraints from M87*}} & \multirow{2}{*}{\textbf{Constraints from SgrA*}}\\
& &   \\

\hline
\multirow{2}{*}{RQCBHs} &$0.6157\leq a\leq0.8511$  & $0<a<0.8066$ \\
&  $0 < \alpha< 0.8985$ & $0 < \alpha< 1.443$  \\
\hline
\multirow{2}{*}{LIRBH-1  \cite{Afrin:2022ztr}}& \multirow{2}{*}{$0 < L_q< 0.0643$} &\multirow{2}{*} {$0 < L_q< 0.0423$} \\
&  &    \\
\hline
\multirow{2}{*}{LIRBH-2  \cite{Afrin:2022ztr}}& \multirow{2}{*}{ $0 < L_q< 0.1253$} & \multirow{2}{*}{$0 < L_q< 0.0821$}   \\
&   &   \\
\hline
\multirow{2}{*}{In Bumblebee gravity \cite{Islam:2024sph}}  &  $-0.0037< \ell < 0.1672  $  & $-0.1162< \ell <0.01123 $   \\
& for $a=0.5$ & for $a=0.5$\\
\hline
\multirow{2}{*}{ Hairy Kiselev black holes \cite{Ahmed:2025zdc}}  & $0.2402 < \l < 1 $ & $0.03304< \l < 0.7415 $  \\
& for $\omega=-2/3$ & for $\omega=-2/3$   \\
\hline
\multirow{2}{*}{Bardeen black holes \cite{KumarWalia:2022aop}} & $ 0\leq g\leq0.30182$ &$0.154 \leq g \leq 0.58367$  \\
& for $a=0.7$ & for  $a=0.2$\\
\hline
\multirow{2}{*}{Hayward blackholes \cite{KumarWalia:2022aop}} &$ 0\leq g\leq0.73627$ & $g\geq 0.4881$\\
& for $a=0.7$   & for $a=0.2$\\
\hline 
\multirow{2}{*}{Ghosh–Culetu black holes
 \cite{KumarWalia:2022aop}} & $0\leq g\leq0.30461$ & $ 0.155 \leq g \leq 0.61116.$ \\
& for $a=0.7$ & for $a=0.2$ \\ 
\hline 
\multirow{2}{*}{Simpson–Visser black holes \cite{KumarWalia:2022aop}}  & \multirow{2}{*}{independent of $g$} & \multirow{2}{*}{independent of $g$} \\
& & \\
\hline

\end{tabular}
\caption{Constraints on deviation parameters of other rotating black hole spacetimes from EHT observations of M87* and Sgr A* and compared with RQCBHs within 1 $\sigma$. The parameters  $a$, $\ell$, $l$, and $g$ are in units of $M$ whereas $\alpha$ is in units of $M^2$. }
\label{bhconst}
\end{table}

\subsection{Comparison with other rotating black hole spacetimes with additional deviation parameter(s)} 
Here, we compare the results of the RQCBHs with other well-motivated rotating black hole spacetimes with additional deviation parameters arising from the modified theories of gravity, including LQG-inspired rotating black holes (LIRBHs) \cite{Afrin:2022ztr}, rotating black holes in Bumblebee gravity \cite{Islam:2024sph}, hairy Kiselev black holes \cite{Ahmed:2025zdc} and regular black holes \cite{KumarWalia:2022aop, Kumar:2020yem, Kumar:2020ltt}.
The Sgr A* and M87* shadows predicted by these models are compared and discerned with the observational constraints provided by the EHT (cf. Table~\ref{bhconst}).

It turns out that the rotating black holes arising from modified theories of gravity are also defined by the metric \ref{metric} with appropriately modified mass function $m(r)$.

\paragraph{LQG-Inspired Rotating Black Holes}
LQG-inspired rotating black holes (LIRBHs) introduce an additional parameter $L_q$, representing quantum effects. Afrin {\it et al.} \citep{Afrin:2022ztr} continued to explore two prominent models of LIRBHs, viz. LIRBH-1 and LIRBH-2. LIRBH-1 is the rotating counterpart \citep{Liu:2020ola} of the semiclassical LQG inspired spherical solution \cite{Modesto:2008im}, while LIRBH-2 is derived using modified NJA \citep{Brahma:2020eos,Yang:2022yvq} with the spherical LQG inspired quantum extension of the Schwarzschild spacetime \citep{Bodendorfer:2019nvy} as a seed metric. Using the EHT results for the shadow diameters of M87* and Sgr A*, the constraints on $L_q$ for LIRBH-1 are $0 < L_q< 0.0643$ and $0 < L_q< 0.0423$, respectively. For LIRBH-2, the constraints are $0 < L_q< 0.1253$ for M87 * and $0 < L_q< 0.0821$ for SgrA * respectively, as summarized in Table~\ref{bhconst}. The predicted values of $L_q$ are also consistent with the EHT observations.

\paragraph{Bumblebee Gravity Black Holes}
Islam {\it et al.} \cite{Islam:2024sph} examined the bumblebee model in which a vector-tensor extension of Einstein-Maxwell theory, Lorentz violation, is introduced via a parameter $\ell$ . This model extends GR by introducing a vector field with a non-vanishing vacuum expectation value, causing spontaneous Lorentz symmetry breaking \citep{Kostelecky:1989jw,Bluhm:2004ep}. The EHT constraints on black holes in bumblebee gravity were extended by \citep{Islam:2024sph,Afrin:2024khy}. The rotating spacetime in Bumblebee Gravity is given by metric (\ref{metric}) with 
\begin{equation}
    M(r) = \frac{M(1+\frac{r\ell}{2M})}{1+\ell}
\end{equation}
By modelling M87* and Sgr A* as rotating black holes in Bumblebee gravity, the shadow angular diameter is examined. For M87 *, within the bound $1\sigma$ of $39\,\mu\text{as} \leq \theta_{sh} \leq 45\,\mu\text{as}$, $\ell$ is constrained to $-0.0037M \leq \ell \leq  0.1672M$ for a spin parameter $a = 0.5M$. For Sgr A*, tighter constraints are obtained: at $a = 0.5M$, $\ell$ ranges from $-0.1162M$ to $ 0.01123M$.

\paragraph{Hairy Kiselev Black Holes}
The rotating hairy Kiselev black hole solution \cite{Ahmed:2025zdc}, derived using the gravitational decoupling approach and modified NJA \citep{Brahma:2020eos,Yang:2022yvq} incorporates a quintessence parameter $\omega$ and a hairy parameter $l$. The EHT constraints on the rotating hairy Kiselev black holes were recently discussed by \cite{Ahmed:2025zdc}.   The rotating hairy Kiselev black hole is given by metric (\ref{metric}) with
\begin{equation} 
M(r) = M-\alpha\frac{r}{2} e^{-r/(M-l/2)}+\frac{1}{2}\frac{1}{r^{3\omega+1}}.
\end{equation}
For different values of $\omega$, the shadow angular diameters of M87* and Sgr A* are used to constrain $l$ \cite{Ahmed:2025zdc}. For  $\omega = -2/3$,  the bounds on $l$ for M87* and SgrA* are  $0.2402M < \l< 1$ and  $0.03304M < \l< 0.7415M$ respectively, as presented in Table~\ref{bhconst}, demonstrating the influence of the hairy and quintessence parameters on the black hole shadow.

Next, we compare our results with four well-known rotating regular black holes, namely Bardeen \citep{Bambi:2013ufa,Kumar:2020yem,Kumar:2020ltt}, Hayward \citep{Bambi:2013ufa,Kumar:2020yem,Kumar:2020ltt}, Ghosh-Culetu \citep{Ghosh:2014pba} and Simpson-Visser black holes \citep{Mazza:2021rgq,Islam:2021ful}. The regularity of these black holes' spacetimes is analysed in terms of the curvature scalars, which are finite everywhere \citep{Walia_2022}. In Ref.~\citep{Kumar:2020yem}, we have shown that the shadows of Bardeen black holes ($g\lesssim 0.26 M$), Hayward black holes ($g\lesssim 0.65 M$), Ghosh-Culetu black holes ($g\lesssim 0.25 M$) and Simpson-Visser (independent of $g$) are indiscernible from Kerr black hole shadows within the current observational uncertainties. Thus, these black holes are strong and viable candidates for astrophysical black holes. Furthermore, Bardeen black holes ( $g\leq 0.30182M$), Hayward black holes ($g\leq 0.73627M$), and Ghosh-Culetu black holes ($g\leq 0.30461M$), within the $1\sigma$ region for $\theta_d= 39\, \mu$as,  are consistent with the observed angular diameter of M87* \citep{Kumar:2020yem}.

\paragraph{ Bardeen spacetime}
The metric (\ref{metric}) characterizing a rotating regular Bardeen black hole \citep{Bambi:2013ufa,Kumar:2020ltt,Kumar:2018ple,Kumar:2020yem,Ghosh:2015pra} is determined using the mass function \citep{Bardeen:1968}
\begin{equation}
M(r)=M\left(\frac{r^2}{r^2 + g^2}\right)^{3/2}, \label{Bardeenmass}
\end{equation}
where $g$ represents the deviation parameter, which can be interpreted as a magnetic monopole charge \citep{AyonBeato:2000zs}. The Kerr black hole is recovered in the limiting case $g\to 0$. Notably, the shadow of the rotating Bardeen black hole ($g\neq0$) exhibits a reduction in size and increased distortion as  $g$ increases, distinguishing it from the Kerr black hole shadow \citep{Walia_2022,Abdujabbarov:2016hnw,Kumar:2020yem,Kumar:2020ltt,Tsukamoto:2014tja}. To constrain $g$ for various values of $a$, the shadow angular diameters of M87* and Sgr A* are exploited to obtain the constraints on $g$ as  $ 0\leq g\leq0.30182M$ for $a=0.7M$ and  $0.154M \leq g \leq 0.58367M$ for  $a=0.2M$ respectively \citep{Kumar:2020ltt,Kumar:2020yem}.

\paragraph{ Hayward spacetime}
Another comprehensively studied regular black hole model was presented by Hayward (\citeyear{Hayward:2005gi}) with additional parameter $\ell$, which determines the length associated with the region concentrating the central energy density, such that modifications in the spacetime metric appear when the curvature scalar becomes comparable with $\ell^{-2}$. The spherically symmetric Hayward black hole model is identified as an exact solution of the general relativity minimally coupled to NED with magnetic charge $g$, where $g$ is related to $\ell$ via $g^3=2M\ell^2$ \citep{Fan:2016hvf}. The rotating Hayward black hole is also described by the metric (\ref{metric}) with the mass function \citep{Hayward:2005gi,Kumar:2018ple}
\begin{equation}
M(r)=\frac{Mr^3}{r^3+g^3}\label{Haywardmass}
\end{equation}
The angular diameter $\theta_{sh} = 39.6192\; \mu$as, which falls within the $1 \sigma$ confidence region with the observed angular diameter of the EHT observation of M87* black hole, strongly constrains the parameter $g$, i.e., $0<g \leq 0.73627M$ for $a=0.7M$ for the Hayward black holes\citep{Kumar:2020yem}. Moreover, the observed angular diameter $\theta_{sh}$ of the SgrA* black hole within $46.9\,\mu\text{as} \leq \theta_{sh} \leq 50\,\mu\text{as}$ places a bound on the parameter $g$ as $g\geq0.4881M$ for $a=0.2M$ \citep{Kumar:2020ltt}.

\paragraph{ Ghosh-Culetu spacetime}
Bardeen and Hayward regular black holes have an asymptotically de Sitter core. The next regular model \citep{Ghosh:2014pba,Culetu:2014lca}, which we hereafter refer to as Ghosh-Culetu black holes, is a novel class of regular black hole with an asymptotic Minkowski core \citep{Simpson:2019mud}. While these regular models share many features with Bardeen and Hayward black holes, there are also notable differences, especially at the deep core \citep{Simpson:2019mud}. The metric was found by Ghosh \citep{Ghosh:2014pba} generalizing the
spherically symmetric regular (i.e. singularity-free) black hole
solution \citep{Culetu:2014lca} to the rotating case with the mass function  \citep{Ghosh:2014pba}
\begin{eqnarray}
M(r)=Me^{-g^2/2Mr},
\end{eqnarray}
where $g$ the NED charge.
The observed angular diameter $\theta_{sh}$ of the M87* black hole, within the $1 \sigma$ confidence level, requires $0\leq g \leq 0.30461M$ for $a=0.7M$ for the Ghosh-Culetu black holes \citep{Kumar:2020yem}. Also, the observed angular diameter $\theta_{sh}$ of the SgrA* black hole within $46.9\,\mu\text{as} \leq \theta_{sh} \leq 50\,\mu\text{as}$ places a bound on the parameter $g$ as $0.155M<g\leq 0.61116M$ for $a=0.2M$ \citep{Kumar:2020ltt}.  

\paragraph{ Simpson-Visser spacetime}
The rotating Simpson-Visser black holes metric is constructed as a modification from the Kerr black hole metric \citep{Shaikh:2021yux,Mazza:2021rgq}, and is given by the (\ref{metric}) with mass function 
\begin{eqnarray}
M(r)=M \sqrt{1+ \frac{g^2}{r^2}}.
\end{eqnarray}
$\Sigma = r^2+g^2 +a^2\cos^2\theta$ and  $\Delta = r^2+g^2+a^2-2 M(r) r$ are different from the previous three black holes. It is important to note that the radial coordinate $r$ is not the same as the areal radius, which is $\sqrt{r^2+g^2}$. The Simpson-Visser spacetime has an interesting feature, namely that the areal radii of the event horizon, the photon sphere, and the shadow are completely independent of parameter $g$, but, the proper distance between these surfaces is $g-$dependent \citep{Lima:2021las}.

The RQCBH model provides a distinct framework for exploring quantum gravity effects, while comparisons with LQG, Bumblebee gravity, hairy Kiselev, Bardeen, Hayward, Ghosh-Culetu and Simpson-Visser black holes reveal the challenges in constraining non-Kerr parameters. The predicted values of $L_q$, $\ell$, $l$ and $g$ are consistent with EHT observations, but distinguishing between these models requires higher precision data. Table~\ref{bhconst} summarizes the constraints on these parameters.

\section{Conclusions}\label{Sec7}
To keep pace with rapidly growing astronomical observations, such as those from the EHT, there is a pressing need for a robust model of RQCBHs. One promising technique uses a modified Newman-Janis algorithm (NJA), a feasible solution-generating method that converts a non-rotating seed metric into a rotating black hole solution. The resulting RQCBH spacetime (\ref{metric}), derived from the seed metric (\ref{metric1}) \cite{Lewandowski:2022zce}, has a relatively simpler form, exhibits intriguing properties, and asymptotically approaches the Kerr solution.
The no-hair theorem establishes that, in GR, black holes are uniquely described by three parameters: mass, spin, and charge \citep{Israel:1967za,Carter:1971zc,Hawking:1971vc}. The Kerr-Newman metric captures this, reducing to the Kerr metric for electrically neutral or astrophysical black holes \citep{Newman:1965my,Kerr:1963ud}. However, actual astrophysical black holes may deviate from this idealized model due to surrounding matter, such as dark matter. In contrast, non-Kerr spacetimes like RQCBHs introduce additional deviation parameters, such as $\alpha$, to account for potential departures from the Kerr metric while reducing it when the deviation parameter vanishes ($\alpha=0$). The quantum correction parameter  $\alpha$ leads to a richer structure of event horizons and black hole shadows. Our study indicates that the QC  parameter $\alpha$ significantly alters the black hole's horizon structure. This can have significant implications for astrophysical processes, particularly energy extraction like the Penrose process, where the expanded ergosphere improves energy extraction capacities. 

This work focuses on studying the RQCBH's shadow and analysing its spacetime properties. Specifically, we demonstrate that the Hamilton-Jacobi equations for this spacetime are separable, leading to null geodesic equations in a first-order differential form.  The available information from the EHT observation of Sgr A* and M87* shadow motivated us to reevaluate the shadow cast by RQCBH while considering the influence of the QC parameter. Notably, RQCBH with parameter $\alpha$ cast larger and more distorted shadows than Kerr black holes; as $\alpha$ increases, the shadow size increases while distortion decreases.  The Shadow observables, such as shadow radius $R_s$ and distortion $\delta_s$, help characterize the size and shape of the shadow and estimate the black hole parameters ($a, \alpha$). Additionally, parameters such as area $A$ and oblateness $D$ are used to determine ($a, \alpha$) explicitly. 

The black shadow observables that describe the shadow shape and size quantify the shadow characteristics. We employed the EHT observables Schwarzschild deviation ($\delta$) and angular diameter ($\theta_{sh}$) of the shadow cast by the SMBHs M87* and Sgr A* by modelling them as RQCBHs to test the viability of the Quantum Gravity at the current observational resolution of the EHT. Our investigation reveals that the EHT observations of M87* and SgrA* lead to constraints on $\alpha$ and $a$. To place constraints on the QC parameter $\alpha$,  we employ the angular size and asymmetry results of the EHT for the M87* black hole and the EHT bounds on the Sgr A* shadow angular diameter and Schwarzschild shadow deviation from the Sgr A* results. For M87*, we have used the calibrated mean shadow diameter, $d_{sh}^{M87*}$ as $37.8~\mu$as. The allowed range of M87* angular shadow diameter is possible for $a \in (0, 0.8511 M)$ at $\alpha=0$  and $a \in (0, 0.6157 M)$ at $\alpha=0.8985 M^2$ for $\theta_o=17$\textdegree and $a \in (0, 0.8262 M)$ at $\alpha=0$  and $a \in (0, 0.9799 M)$ at $\alpha=0.4141 M^2$ for $\theta_o=90$\textdegree. Likewise, for SgrA*, we have used the mean shadow diameter, $d_{sh}^{Sgr A*}$ as $48.7~\mu$as which strongly constrains the parameters, such that $0.8066 M \leq a \leq a_{c}$ and  $0<\alpha<1.443M^2$  at $\theta_o=50$\textdegree. At $\theta_o=90$\textdegree, $0.894 M \leq a \leq a_{c}$ and  $0<\alpha<1.447M^2$. Here, $a_{c}$ is the critical value of parameter $a$. Further, modelling M87* and Sgr A* as  RQCBHs, the entire parameter space is satisfied for both for the inferred  EHT bounds. In conjunction with the EHT constraints for M87* and Sgr A*, our analysis indicates an agreement of a significant region of parameter space of RQCBHs with the EHT observations. It is possible, therefore, to assert that the RQCBHs could be reliable candidates for astrophysical black holes.
Finally, we employ two well-known parameter estimation techniques to calculate the parameters of RQCBH viz. $\alpha$ and $a$.

We compared the results of RQCBHs with other well-motivated rotating black hole spacetimes incorporating deviation parameters from modified theories of gravity, such as LIRBHs \cite{Afrin:2022ztr}, Bumblebee gravity black holes \cite{Islam:2024sph}, hairy Kiselev black holes \cite{Ahmed:2025zdc}, and regular black holes \cite{KumarWalia:2022aop, Kumar:2020yem, Kumar:2020ltt}. The predicted shadows of Sgr A* and M87* in these models were analyzed using the observational constraints provided by the EHT (cf. Table~\ref{bhconst}).
Our analysis showed that while some alternative black hole models exhibited noticeable deviations from the Kerr shadow, many remained consistent with EHT constraints, making distinguishing them from current observational data challenging. The RQCBH model provided a unique framework for exploring quantum gravity effects in black hole imaging. Comparisons with the models mentioned above revealed how different modifications of the mass function $m(r)$ influenced the black hole observables. The parameters $L_q$, $\ell$, $l$, and $g$ in the models mentioned above were within the limits set by the EHT. However, the inability to distinguish between these models conclusively underscores the need for higher-precision observations.
Future high-resolution imaging, particularly with the next-generation EHT (ngEHT), combined with complementary observations such as gravitational waves and X-ray spectroscopy, may provide stringent constraints and help distinguish between different black hole solutions. Our results indicate that RQCBHs serve as viable quantum-corrected alternatives to Kerr black holes, highlighting the necessity for further observational and theoretical investigations into black holes beyond GR.

Thus, from our perspective, distinguishing an RQCBH from a Kerr black hole by observing black hole shadows is challenging. In the meantime, the upcoming ngEHT and space-bas submillimeter interferometry technologies promise to deliver higher-resolution pictures of black holes.  Our results emphasize the potential for using EHT observations to place constraints on the parameters of RQCBH models, presenting useful understanding into the effects of QC on astrophysical black holes.

\section*{Acknowledgements}
S.G.G. gratefully acknowledges the support from SERB-DST through project No. CRG/2021/005771. S.U.I thank the University of KwaZulu-Natal and the NRF for the postdoctoral research fellowship.

\section*{Data Availability}

We have not generated any original data in the due course of this study, nor has any third-party data been analysed in this article.

\bibliographystyle{apsrev4-1}
\bibliography{RQCBH}
\end{document}